\documentclass[aps,prd,twocolumn,a4paper,superscriptaddress]{revtex4-1}
\usepackage{graphicx}
\usepackage{multirow}
\usepackage{latexsym}
\usepackage{hyperref}
\usepackage[british]{babel}
\usepackage{amsmath}

\begin{document}

\title{Extending and Calibrating the Velocity dependent One-Scale model for Cosmic Strings with One Thousand Field Theory Simulations}
\author{J. R. C. C. C. Correia}
\email{Jose.Correia@astro.up.pt}
\affiliation{Centro de Astrof\'{\i}sica da Universidade do Porto, Rua das Estrelas, 4150-762 Porto, Portugal}
\affiliation{Instituto de Astrof\'{\i}sica e Ci\^encias do Espa\c co, Universidade do Porto, Rua das Estrelas, 4150-762 Porto, Portugal}
\affiliation{Faculdade de Ci\^encias, Universidade do Porto, Rua do Campo Alegre 687, 4169-007 Porto, Portugal}
\author{C. J. A. P. Martins}
\email{Carlos.Martins@astro.up.pt}
\affiliation{Centro de Astrof\'{\i}sica da Universidade do Porto, Rua das Estrelas, 4150-762 Porto, Portugal}
\affiliation{Instituto de Astrof\'{\i}sica e Ci\^encias do Espa\c co, Universidade do Porto, Rua das Estrelas, 4150-762 Porto, Portugal}

\date{27 June 2019}

\begin{abstract}
Understanding the evolution and cosmological consequences of topological defect networks requires a combination of analytic modeling and numerical simulations. The canonical analytic model for defect network evolution is the Velocity-dependent One-Scale (VOS) model. For the case of cosmic strings, this has so far been calibrated using small numbers of Goto-Nambu and field theory simulations, in the radiation and matter eras, as well as in Minkowski spacetime. But the model is only as good as the available simulations, and it should be extended as further simulations become available. In previous work we presented a General Purpose Graphics Processing Unit implementation of the evolution of cosmological domain wall networks, and used it to obtain an improved VOS model for domain walls. Here we continue this effort, exploiting a more recent analogous code for local Abelian-Higgs string networks. The significant gains in speed afforded by this code enabled us to carry out 1032 field theory simulations of $512^3$ size, with 43 different expansion rates. This detailed exploration of the effects of the expansion rate on the network properties in turn enables a statistical separation of various dynamical processes affecting the evolution of the network. We thus extend and accurately calibrate the VOS model for cosmic strings, including separate terms for energy losses due to loop production and scalar/gauge radiation. By comparing this newly calibrated VOS model with the analogous one for domain walls we quantitatively show that energy loss mechanisms are different for the two types of defects.
\end{abstract}
\maketitle
\allowdisplaybreaks

\section{\label{intr}Introduction}

The breaking of some large symmetry, presumably underpinning a Grand Unified Theory (GUT), is thought to have occurred during the early stages of the evolution of the Universe. One possible byproduct of such a phase transition is the production of topological defects by means of the Kibble mechanism \cite{Kibble:1976sj}. One type of defect that is generally benign (\textit{i.e.,} not expected to dominate the energy density of the Universe) is produced when an axial symmetry is broken. These are known as cosmic strings. They are relic byproducts in many GUT-scenarios, such as supersymmetric GUTs \cite{Jeannerot:2003qv}, and can even be produced in brane inflation \cite{Sarangi:2002yt}. In the latter case they are dubbed instead cosmic superstrings; for a review on these, see \cite{Polchinski:2004ia}.

Networks of cosmic strings typically evolve toward a scaling (more accurately, scale-invariant) regime, in which the average string separation grows in proportion to horizon size and the network-averaged velocity squared is constant \cite{Book}. The dynamics of the defects, subsequent energy losses and network evolution can greatly impact observational signatures (most notably in the cosmic microwave and stochastic gravitational wave backgrounds), given that these are seeded by the energy-momentum tensor of the defects themselves. As such, understanding what processes will predominantly drive network evolution and how energy is transferred from large to small scales is important for current \cite{LIGODefects,PlanckDefects} as well as future constraints \cite{LISA,CORE}. It is thus unfortunate that for decades the two main types of cosmic string simulations (field theory Abelian Higgs \cite{Moore:2001px,Bevis:2006mj} and Goto-Nambu connected segments \cite{BB,AS,FRAC,VVO}) largely disagree in terms of loop production (a key energy loss mechanism), the amount of small-scale structure on the strings \cite{Hindmarsh:2008dw,Blanco} and the resulting large-scale properties of the network \cite{Moore:2001px,Hindmarsh:2017qff}.  

Numerical simulations of defect networks need to be complemented by analytic models. The canonical one is the Velocity-dependent One-Scale (VOS) model, first introduced for cosmic strings \cite{Martins:1996jp,MS2} and then subsequently developed for several other types of defects---for a recent overview, see \cite{Book}. The obvious advantages of an analytic model are somewhat offset by the fact that any such model must be calibrated by numerical simulations, and therefore the model can only be as useful (and reliable) as the available simulations. In the case of cosmic strings, the VOS model has so far only been calibrated using a small number (around 10 to 20) of field theory and Goto-Nambu simulations \cite{Moore:2001px,Moore}. But this analytic modeling can---and should---always be improved, as further computational resources become available.

For the simpler case of domain walls, an improved and accurately calibrated VOS model has been recently developed \cite{Rybak1,Rybak2}. To a large extent this has relied on the field theory simulation of a large number of different expansion rates, as opposed to just the 'canonical' cases of the radiation and matter era (as well as the more simplistic and unrepresentative case of Minkowski spacetime). This is important because the multiple different expansion rates (which \textit{inter alia}, affect the defect velocities) allow one to quantitatively identify the velocity dependencies of various physical processes contributing to the network dynamics (at least in a statistical sense) and thereby to separately include them in the VOS model. Specifically, for domain walls the improved VOS model includes separate energy loss terms from both scalar radiation and from chopping of wall blobs (the analogues of cosmic string loops), with the former clearly being the dominant one. In parallel, a General Purpose Graphics Processing Unit (GPU) implementation of the evolution of cosmological domain wall networks has also been developed \cite{PhysRevE.96.043310}, and its results agree with previous codes \cite{Biases}.

In the present work we bring together the two developments described in the previous paragraph, in the context of the more complex but also more interesting case of cosmic strings. Specifically, we use a recently developed GPU-accelerated field theory simulation of Abelian-Higgs cosmic strings, described and validated in \cite{Correia:2018gew}, to extend and calibrate the cosmic strings VOS for expansion rates in which the strings are relativistic when scaling. Compared to the small number of simulations available in earlier work, our latest codes and available hardware allow us to perform a total of 1032 simulations (each with a $512^3$ box size) for a total of 43 different expansion rates (radiation, matter and 41 other rates), all in an eminently reasonable amount of time. A comparison with the domain walls case also produces some important differences. Specifically, comparing the newly calibrated strings VOS model with the analogous one for domain walls, we quantitatively show how the energy loss mechanisms differ in the two cases. The observational consequences of these differences will be explored in future work.

The plan of the rest of the paper is as follows. We start in Sect. \ref{code} with a brief outline of the simulation code. The outputs of each simulation, which are used as diagnostics of the evolution of the network (in particular for identifying when it has reached scaling) and will be subsequently used to calibrate the model, are described in Sect. \ref{outputs}. In Sect. \ref{evos} we briefly introduce the current version of the VOS model and motivate and describe an extended version thereof. This extended model is then calibrated, using the aforementioned simulations, in Sect. \ref{calib}, which also includes a comparison of the newly calibrated strings VOS model with the analogous one for domain walls. Finally we summarize our results and highlight some potential impacts in Sect. \ref{concl}.

\section{\label{code}Simulation code}

We start with a brief description of our GPU-accelerated simulation code for Abelian-Higgs cosmic strings. A somewhat more technical description of the code, including a discussion of its performance and validation, can be found in \cite{Correia:2018gew}. We note that all our simulations will be done in expanding universes, with the scale factor evolving as a power law $m$ of physical time $t$, that is $a\propto t^m$. We will discuss our choices of $m$ below and in the following sections.

Abelian-Higgs strings arise as topologically stable solutions of the equations of motion of the following $U(1)$ locally invariant Lagrangian density,
\begin{equation}
\mathcal{L}=|D_\mu \phi|^2 - \frac{\lambda}{4}(|\phi|^2 -\sigma^2)^2 - \frac{1}{4e^2}F^{\mu \nu}F_{\mu \nu}
\end{equation}
where $\phi$ is a complex scalar field, the electromagnetic field tensor is given by $F_{\mu \nu} = \partial_\mu A_\nu - \partial_\nu A_\mu$, $A_\mu$ is the gauge field (where the gauge coupling $e$ has been absorbed), $D_\mu \phi$ is the gauge covariant derivative given by $D_\mu = \partial_\mu -iA_\mu$ and $\lambda$ and $e$ are coupling constants. Through standard variational techniques, and under the assumptions of both the temporal gauge ($A_0 =0$) and a Friedmann-Lemaitre-Robertson-Walker background ($g_{\mu \nu} = a^2 diag(-1,1,1,1)$), one obtains the following equations of motion
\begin{equation}
\ddot{\phi} + 2\frac{\dot{a}}{a}\dot{\phi} = D^jD_j\phi -\frac{a^{2}\lambda}{2} (|\phi|^2 - \sigma^2) 
\end{equation}
\begin{equation}
\dot{F}_{0j} = \partial_j F_{ij} -2a^2 e^2 Im[\phi^* D_j \phi]
\end{equation}
along with Gauss's law,
\begin{equation}
\partial_i F_{0i} = 2 a^2 e^2 Im[\phi^* \dot{\phi}]\,,
\end{equation}
where $\dot{a}$ denotes the derivative of the scale factor with respect to conformal time. 

These equations of motion have to be modified, as shown in \cite{PRS,Bevis:2006mj}, in order to ensure that the string radius does not fall below the lattice spacing in the simulations and that Gauss's law is still preserved (independent of the modifications to string radius). To do so, the variables which describe the Compton wavelengths, $\lambda$ and $e$, must vary as
\begin{align}
\lambda = \lambda_0 a^{2(1-s)} && e = e_0 a^{(1-s)} \,,
\end{align}
where the parameter $s$ controls how the co-moving string radius evolves over time, as was done by \cite{Bevis:2006mj}, with $s=0$ corresponding to constant comoving width (known as the Press-Ryden-Spergel algorithm \cite{PRS}) and positive values of $s$ corresponding to shrinking comoving string radius. Inserting these into the action and performing variation then yields, as \cite{Bevis:2006mj} has shown,
\begin{equation}
\ddot{\phi} + 2\frac{\dot{a}}{a}\dot{\phi} = D^jD_j\phi -\frac{a^{2s}\lambda_0}{2} (|\phi|^2 - \sigma^2) 
\end{equation}

\begin{equation}
\dot{F}_{0j} + 2(1-s)\frac{\dot{a}}{a}F_{0j} = \partial_j F_{ij} -2a^{2s} e_0^2 Im[\phi^* D_j \phi]\,.
\end{equation}
Note that the extra second term in the second equation (left-hand side) is responsible for ensuring Gauss's law preservation at $s\neq1$.

To find a suitable discretization scheme, we must turn to the principles of lattice gauge theory \cite{Wilson:1974sk}, which allow us to write the gauge on the lattice as the link operator
\begin{equation}
U_j^x = e^{-i A_j}\,,
\end{equation}
defined half-way (at links) between lattice points spaced by $\Delta x$. Note that these links are then technically at sites $x+1/2k_j$, however we re-labeled them to lie at $x$ for convenience. In the above definition we have re-scaled the gauge field as $A'_j \rightarrow A_j \Delta x $ which in turn implies that the electric field $E_j = F_{0j}$ is re-scaled in the same way, as it is merely the time derivative of the gauge field. The scalar fields will reside at lattice sites. Going around a lattice square of size $\Delta x^2$ we can write the following product of link variables, $\Xi_{ij}$
\begin{equation}
\begin{split}
\Xi_{ij} &= U_j^x U_i^{x+k_j}  (U_j^{x+k_i})^* (U_i^{x})^* \\
&= exp[i \Delta x (\partial^+_i A'_j(x) -\partial^+_j A'_i(x) )]\,,
\end{split}
\end{equation}
which is known as the plaquette operator. Here the electromagnetic field tensor is already apparent. From this, we can subsequently write down the gauge field strength
\begin{equation}
\frac{1}{2}F_{ij}F_{ij} = \sum_i \sum_j \frac{1}{\Delta x^4} \bigg(1-Re[\Xi_{ij}]\bigg)\,.
\end{equation}
For convenience, we will also define the backward derivative of $F_{ij}$,
\begin{equation}
\partial^-_j F_{ij} =  \frac{1}{\Delta x^3}\sum_{j\neq i}Im[\Xi_{ij}(x)] - Im[\Xi_{ij}(x-k_j)]\,.
\end{equation}
In addition, we can then define (forward) gauge covariant derivatives
\begin{equation}
D_j^+\phi^x = \frac{1}{\Delta x} \left[U_j^x\phi^{x+k_j} - \phi^{x}\right]\,,
\end{equation}
and subsequently a Laplacian stencil
\begin{equation}
D_j^-D_j^+\phi^x = \sum_j \frac{1}{\Delta x^2} [ U_j^x\phi^{x+k_j} - 2\phi_j^{x} +(U_j^{x-k_j})^* \phi^{x-k_j}]\,.
\end{equation}

We now have all the ingredients to recover the lattice discretization of \cite{Bevis:2006mj}. One needs to take the equations of motion and create the following staggered leapfrog (second order in time) evolution scheme
\begin{equation}
\begin{split}
(a^2\Pi)^{x,\eta+\frac{1}{2}} &= (a^2\Pi)^{x,\eta-\frac{1}{2}}
+\Delta\eta a_\eta^2 [D_j^-D_j^+\phi^{x,\eta}\\ &- \frac{\lambda_0}{2} a_\eta^{2s}(|\phi^{x,\eta}|^2-\sigma^2)\phi^{x,\eta}]
\end{split}
\end{equation}
\begin{equation}
\begin{split}
\bigg(\frac{E_i}{e^2}\bigg)^{x,\eta+\frac{1}{2}} &= \bigg(\frac{E_i}{e^2}\bigg)^{x,\eta-\frac{1}{2}}
+\frac{\Delta x \Delta\eta}{e_\eta^2} [-\partial_j^- F_{ij}\\ &+2e_0^{2}a_\eta^{2s}Im[\phi^* D_i^+ \phi]^{x,\eta}]
\end{split}
\end{equation}
\begin{equation}
\phi^{x,\eta+1} = \phi^{x,\eta} + \Delta \Pi^{x,\eta+\frac{1}{2}}
\end{equation}
\begin{equation}
A^{x,\eta+1}_i = A^{x,\eta}_i + \Delta E^{x,\eta+\frac{1}{2}}_i\,,
\end{equation}
where we have dropped the prime superscript to indicate that $A$ and $E$ are re-scaled, and summation over index j is implicit. Since this is a leapfrog scheme, fields $E_i$ and $\Pi$ (the time derivatives of $A_i$ and $\phi$) are evaluated at half-timesteps of conformal time $\eta\pm 1/2$ and used to evolve $\phi$ and $A_i$ one full step $\eta \rightarrow \eta+1$.

Gauss's law takes the form
\begin{equation}
G = \partial_j^-E_i - 2e_0^2a^{2s}Im[\phi^{x,\eta,*}\Pi^{x,\eta-\frac{1}{2}}] = 0
\end{equation}
to order $\mathcal{O}(\Delta x^2$) and $\mathcal{O}(\Delta \eta^2$). In the work that follows, we will use a lattice spacing of $\Delta x = 0.5$, a timestep size of $\Delta \eta = 0.1$, couplings $\lambda_0=2$ and $e_0=1$ (\textit{i.e.,} the Bogmolny'i limit) and a symmetry breaking scale $\sigma=1$. All simulations start with an initial conformal time $\eta_0=1$ and evolve until the horizon reaches one half of the box size. In the continuum limit (ie. when lattice spacing $\Delta x$ vanishes), this evolution scheme reduces to the above equations of motion. 

We choose to keep $s=0$ in order to avoid having to tune a possible core growth phase (determining a reasonable choice of negative $s$ and for how long in conformal time such a phase should optimally last) at each expansion rate. Such a numerical trick is common in previous field theory simulations, but it is of limited usefulness for our present purposes, which rely on a comparison of multiple expansion rates. In addition, simulations with core growth and then realistic shrinking require more time to reach scaling than analogous simulations with constant co-moving width, which again would make them detrimental to our main goal of calibrating the VOS model. We note that a comparison between the $s=0$ case and the physically correct $s=1$ case has been done, for the radiation era, in our previous work validating the code \cite{Correia:2018gew} and the results of both simulations are the same within one-sigma statistical uncertainties. Thus $s=0$ correctly reproduces the dynamics of a string network, at least to the level of precision required for our analysis.

Last but not least, we do not facilitate the network's relaxation to the scaling regime, \textit{e.g.} by including in the simulations a transient period of gradient flow to dampen the network (as is sometimes done in the literature), but always evolve the correct discretized equations. Indeed this is the main conceptual difference (and novelty) of our work with respect to previous approaches. Previous studies relying on simulations are statistics-limited (ie, the results come from only a few simulations in the radiation or matter era). Since one is mainly interested in the properties of the network in the scaling regime, one is then compelled to artificially accelerate the approach to scaling, and a consequence of introducing this fake damping is that the radiation in the box is lost. In our case, as shown later in the paper, we have a model which can separately account for and distinguish (in an averaged, statistical sense) the energy in the defects and in radiation. Thus in our approach radiation in the box is not a problem---on the contrary, it's an unexplored opportunity, and artificially removing it would mean losing information that is crucial for the modeling. Having more than one thousand simulations means that we are not statistics-limited, and are able to calibrate an improved 6-parameter VOS model with reasonable uncertainties on the model parameters. In other words, our approach leads to a VOS model calibration that might be slightly less precise than it would have been if radiation were artificially removed, but also to one that is more accurate.

There is a practical problem with this discretization, particularly when evolving the simulations at relatively large expansion rates ($m\ge0.9$ at single precision): the divisions and multiplications of $a^2$ factors can, at early time-steps, go beyond the scope of one's precision and thus result in field variables being evaluated to NaN. To avoid this we can re-write the top two equations as
\begin{equation}
\begin{split}
(1+\delta)\Pi^{x,\eta+\frac{1}{2}} &= (1-\delta)\Pi^{x,\eta-\frac{1}{2}}+\Delta\eta  [D_j^-D_j^+\phi^{x,\eta} \\ &-\frac{\lambda_0}{2} a_\eta^{2s}(|\phi^{x,\eta}|^2-\sigma^2)\phi^{x,\eta}]
\end{split}
\end{equation}
\begin{equation}
\begin{split}
(1+\omega)E^{x,\eta+\frac{1}{2}}_i &=  (1-\omega)E^{x,\eta-\frac{1}{2}}_i +\Delta\eta [-\partial_i^- F_{ij} \\
&+ 2e_0^{2}a^{2s}_\eta Im[\phi^* D_i^+ \phi]^{x,\eta} ]
\end{split} 
\end{equation}
where
\begin{equation}
\omega=\delta(1-s)
\end{equation}
\begin{equation}
\delta=\frac{1}{2} \alpha \frac{dlna}{dln\eta}\frac{\Delta \eta}{\eta} = \frac{1}{2} \alpha \frac{m \Delta \eta}{(1-m)\eta}\,.
\end{equation} 
Note that $\delta$ is responsible for Hubble damping the scalar field and $\omega$ is responsible for damping the gauge field.

\begin{figure}
\includegraphics[width=\columnwidth]{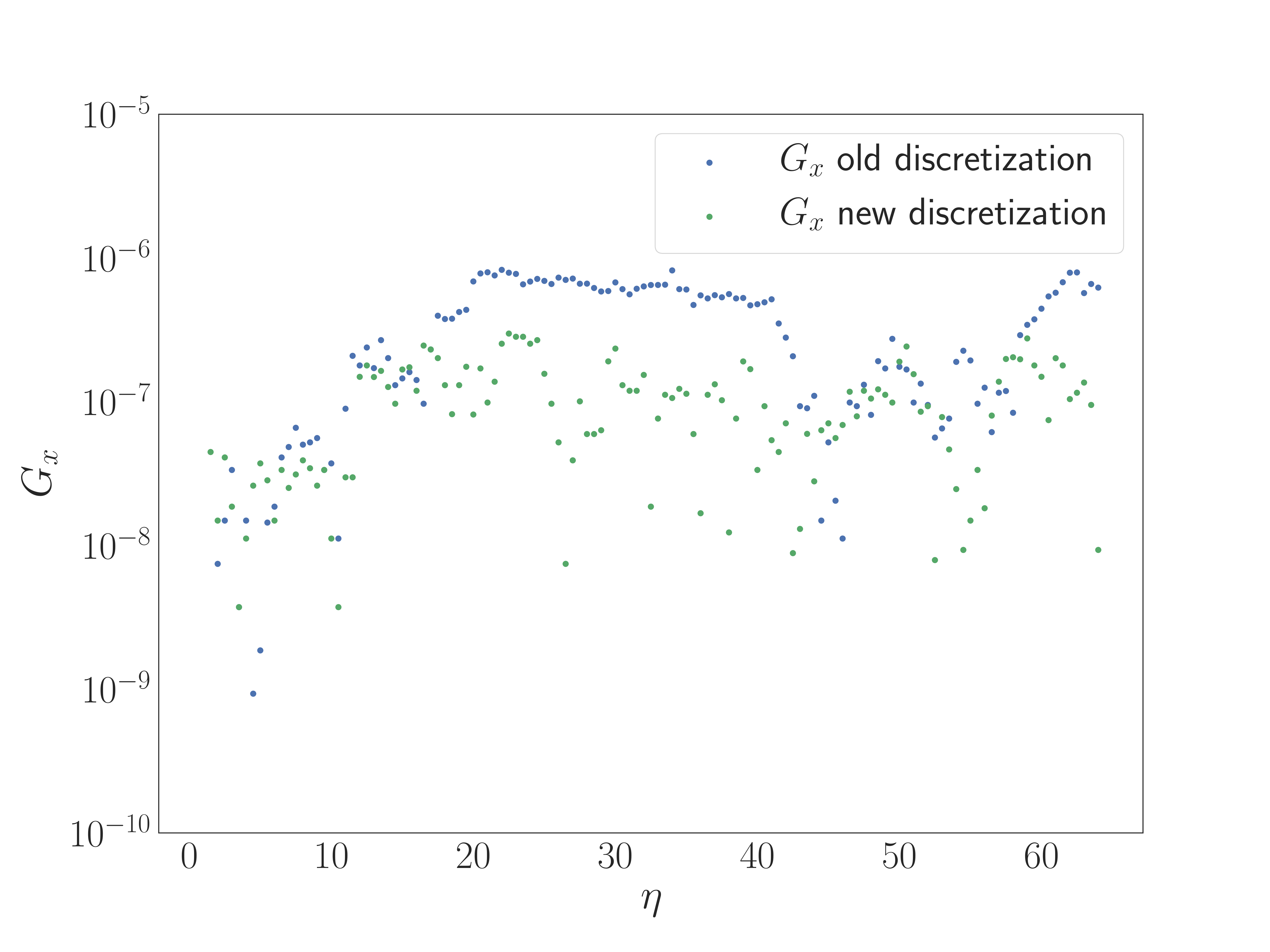}
\includegraphics[width=\columnwidth]{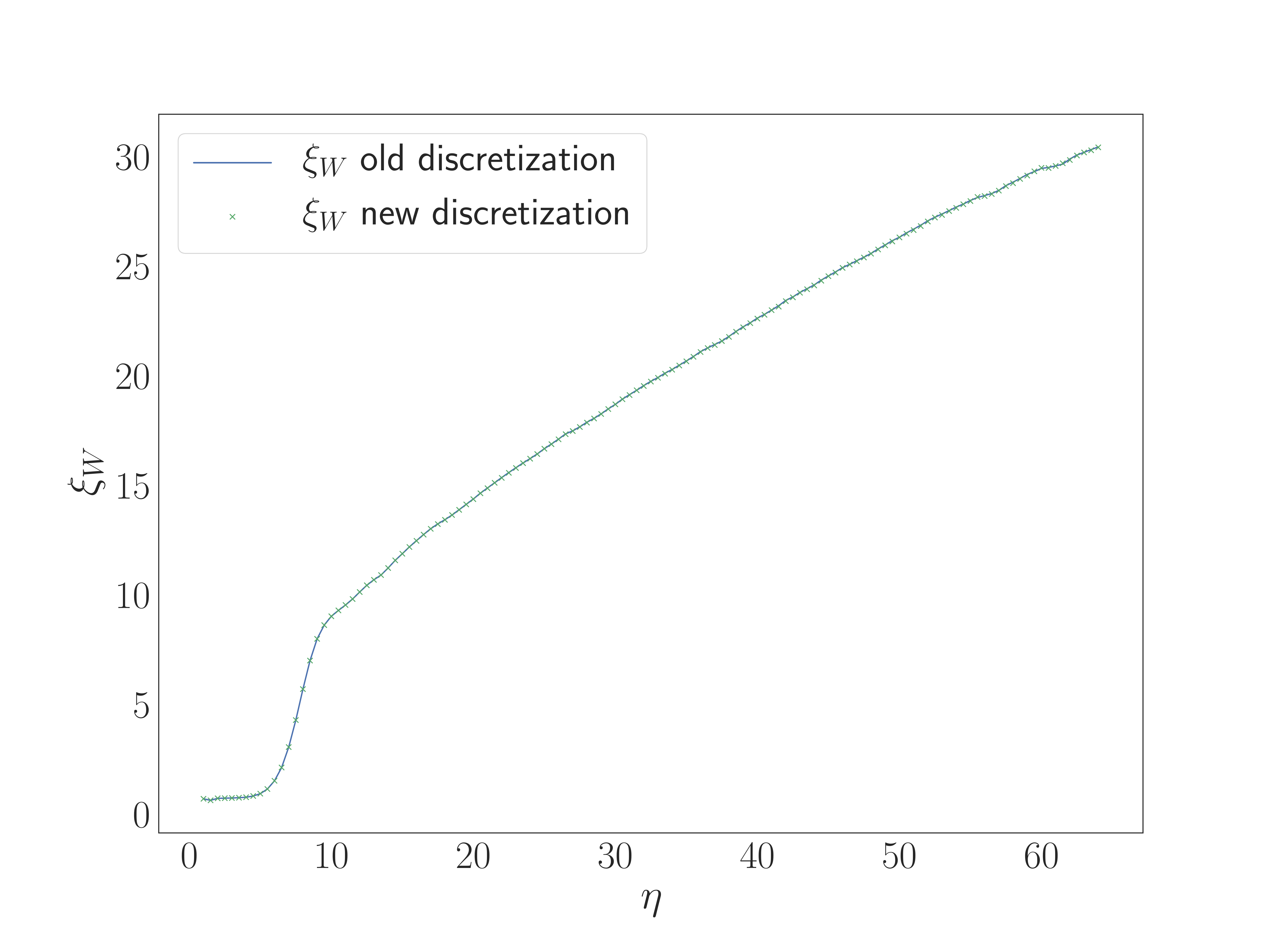}
\caption{The top panel shows the Gauss's law violation operator $G_x$ at lattice site $i,j,k=0,0,0$ at single precision for a box of size $256^3$, while the bottom panel shows a winding based correlation length estimator $\xi_W$ for two simulations using the same initial conditions, with either the new or the old discretizations, described in the text. For this comparison we use the same parameters as in the rest of the paper: $\Delta x = 0.5$, $\Delta \eta = 0.1$, $\lambda_0=2$, $e_0=1$ and $\sigma=1$.\label{fig1}}
\end{figure}

This is similar to the discretization for walls seen in \cite{PRS}, since the scheme is now Crank-Nicolson at the first order with respect to the time terms. Note that our previous problem with the physical string thickness is solved by selecting $s=0$ (such that $a^{2s}$ is replaced by $1$) and $\delta$ and $\omega$  are computed directly from the expansion rate. We set $\alpha=2.0$ so as to recover the equations of motion in the continuum limit. Extensive testing shows that this evolution scheme preserves Gauss's law and reproduces the dynamics of the network up to $0.02\%$ at worst. For a brief comparison at $256^3$ box size, see Fig. \ref{fig1}, in which Gauss's law violations at single precision are displayed in the top panel, together with the behavior of a winding based correlation length estimator (see the next section for a description of the estimator itself).

Within their statistical uncertainties, and for the radiation and matter era (the only cases where such a comparison is possible), our results are consistent with the measured values of  \cite{Bevis:2006mj} for the same timestep range (for a comparison see \cite{Correia:2018gew}), and also with the ones that can visually inferred from Figure 8 of  \cite{Hindmarsh:2017qff}, which has values directly measured in their simulations. Our results are not compatible with the ones listed in Table IX of  \cite{Hindmarsh:2017qff} but we emphasize that those correspond to extrapolations to zero radii strings, and would therefore be appropriate for comparison with Goto-Nambu simulations but not with our work.

\section{\label{outputs}Simulation diagnostics}

For the purpose of calibrating the VOS model, the two essential diagnostics that must be extracted from the simulations are a correlation length $\xi$ and a root mean squared velocity $<v^2>$. Before describing how to compute these outputs in the simulations, we must first define some relevant quantities. First, the Lagrangian density
\begin{equation}
\begin{split}
\mathcal{L}_x &= \frac{1}{2e^2  a^2  \Delta x^2 } E^2  -   \frac{1}{2e^2  a^2  \Delta x^4 }\sum_i \sum_j \bigg(1-Re[\Xi_{ij}]\bigg)  \\
& +|\Pi|^2 -|D^+\phi|^2 - a^2 V(\phi) \\
&= E - B +P -D -V\,,
\end{split}
\end{equation}
where for convenience in the last line we have also introduced a simplified notation for each of its components. From here we can also define an energy density and a pressure,
\begin{equation}
\rho_x = E + B + P + D + V
\end{equation}
\begin{equation}
p_x =\frac{E + B}{3} + P - \frac{1}{3}D - V\,.
\end{equation}

\begin{figure*}
\includegraphics[width=1.0\columnwidth]{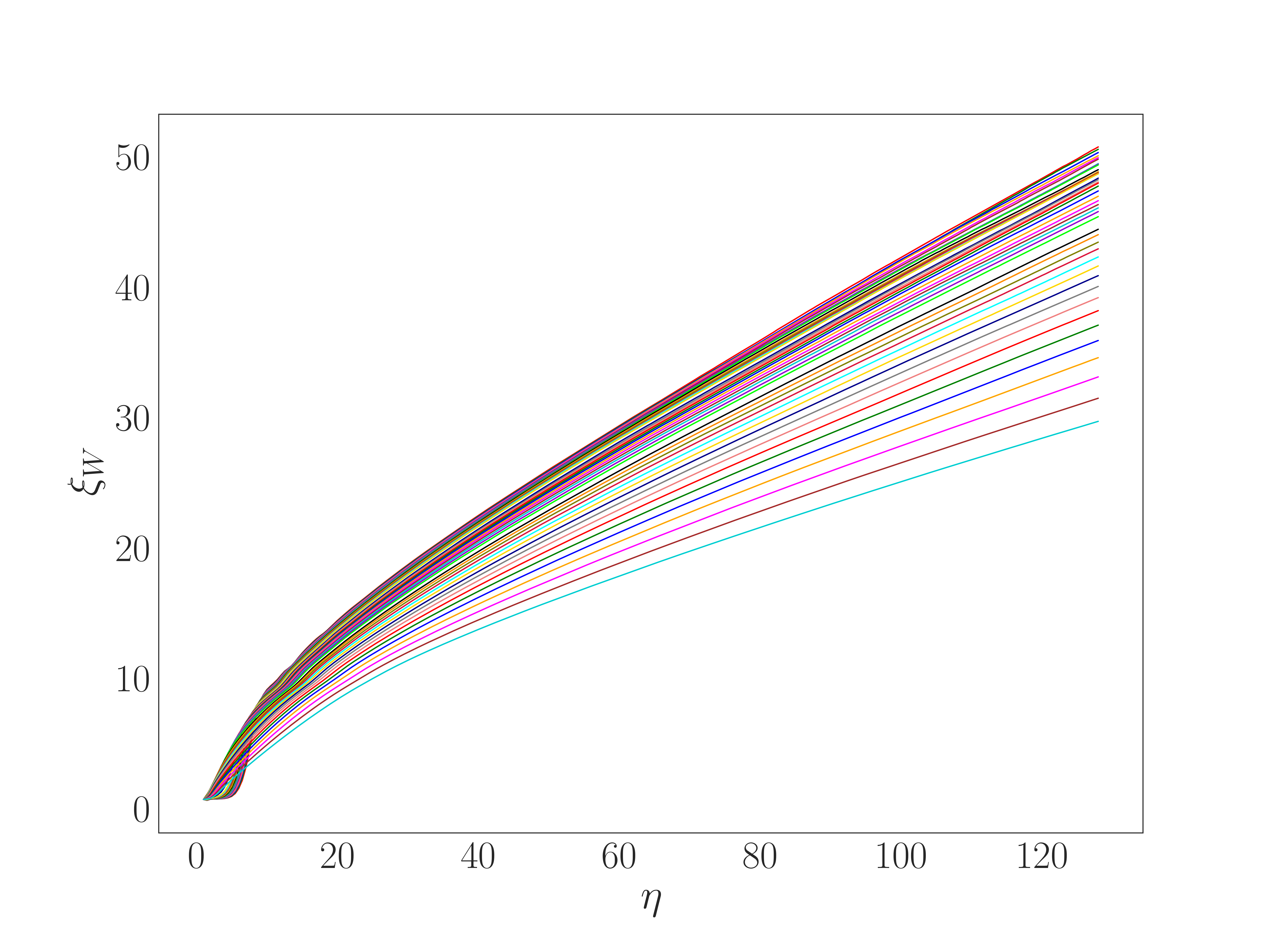}
\includegraphics[width=1.0\columnwidth]{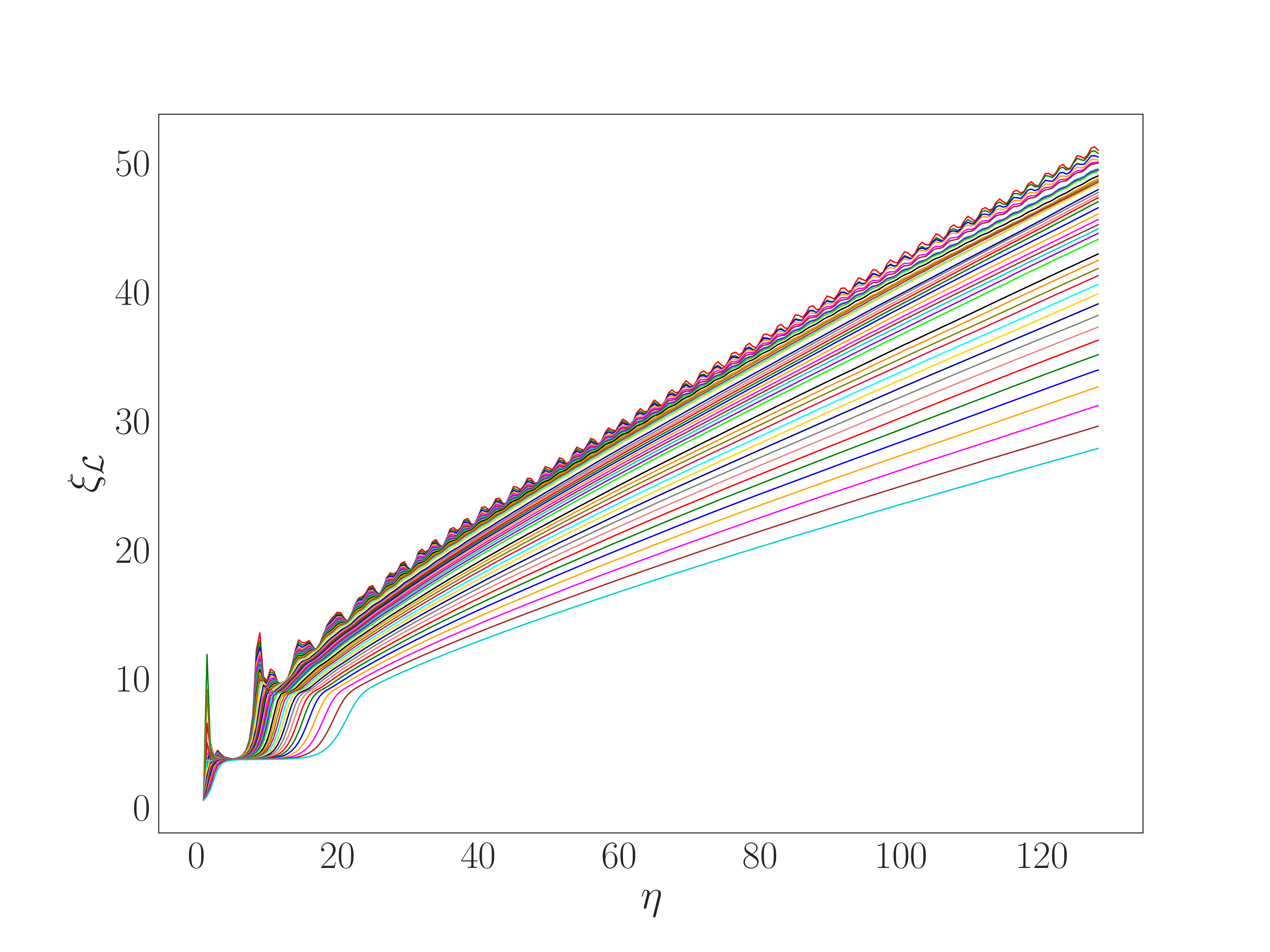}
\includegraphics[width=1.0\columnwidth]{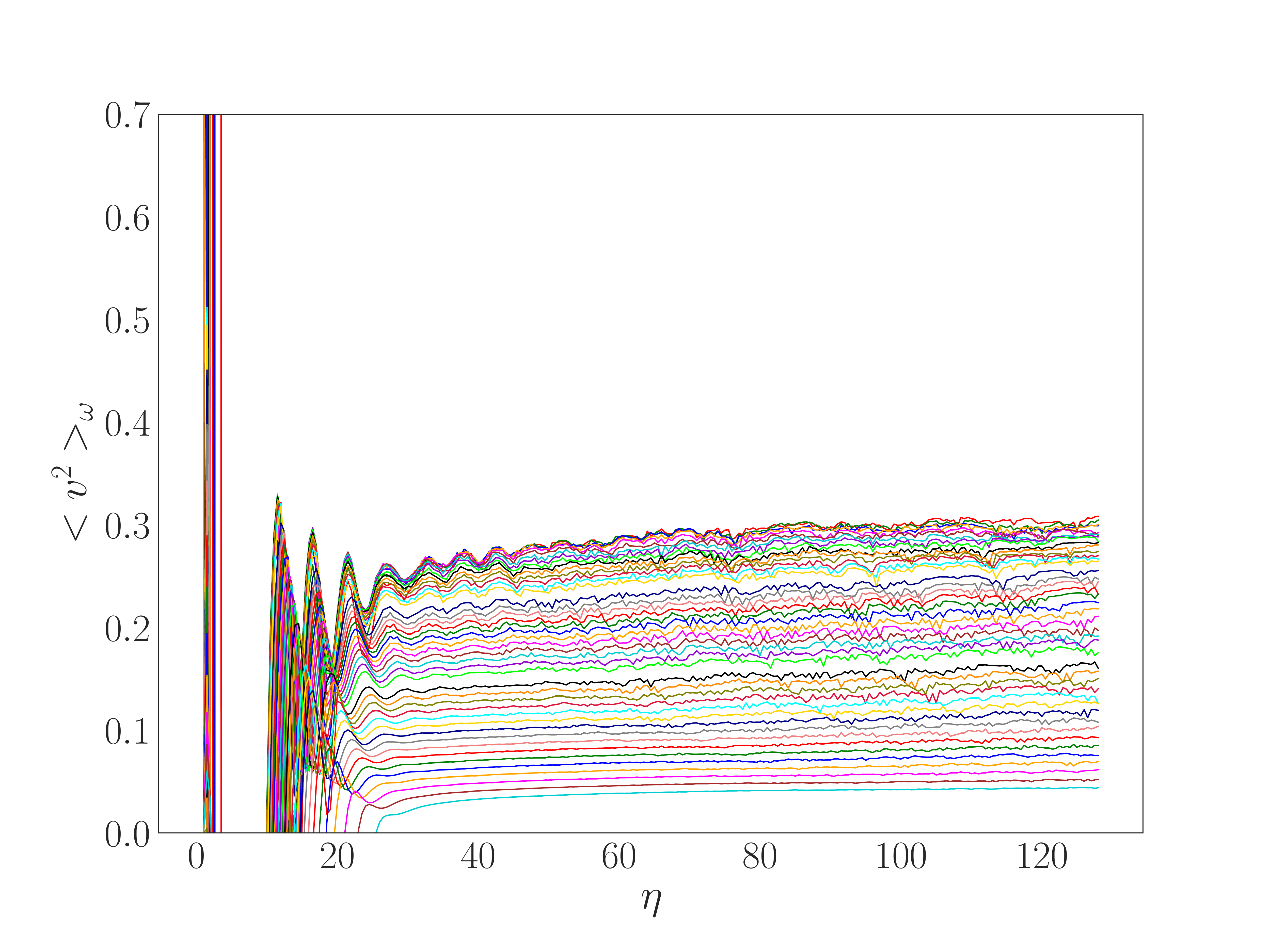}
\includegraphics[width=1.0\columnwidth]{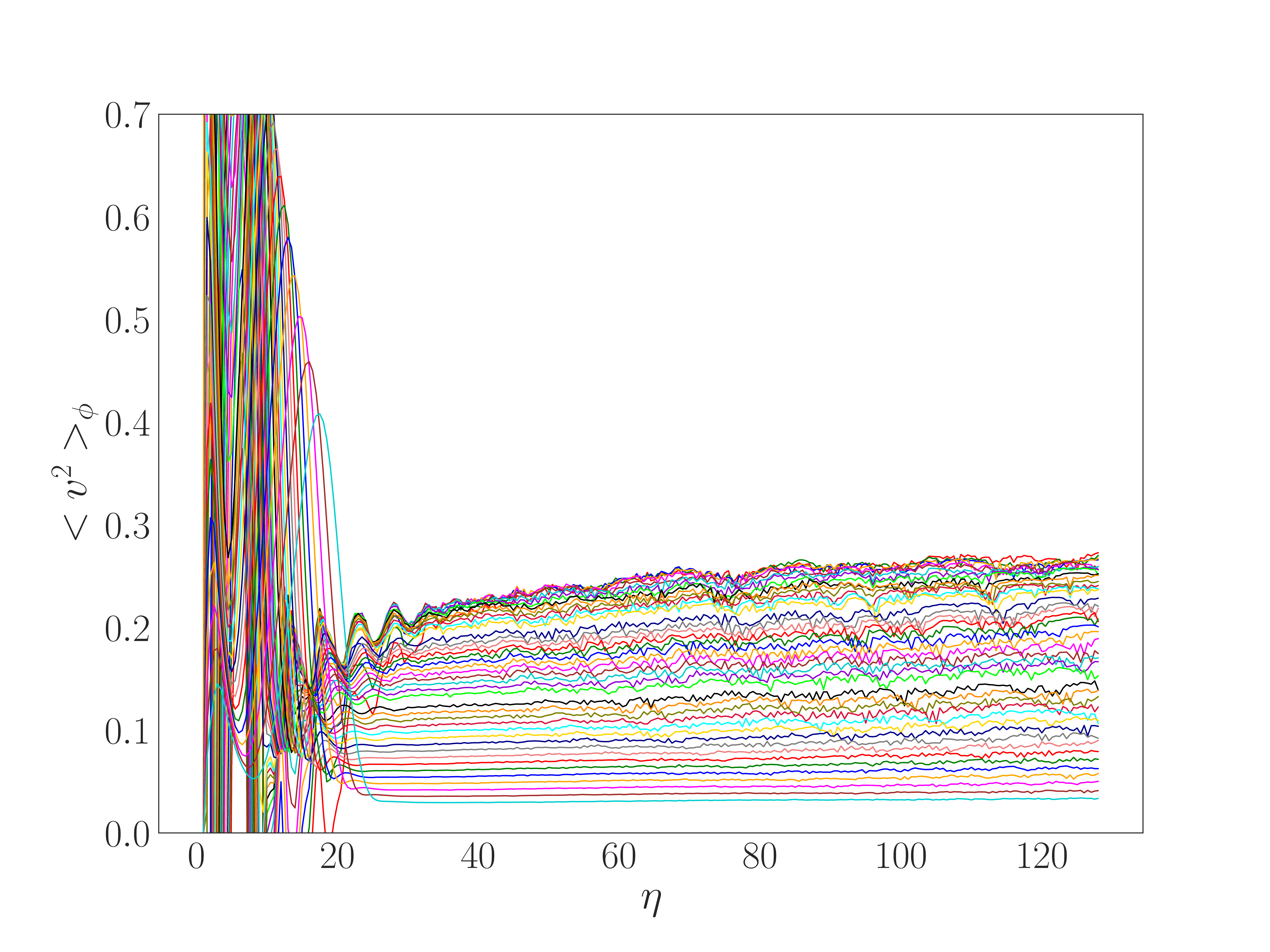}
\caption{The average evolution of the correlation length $\xi$ (top panels) and the mean average velocity $<v^2>$  (bottom panels) for sets of 12 runs at each expansion rate $m$ in the range $[0.5,0.95]$. The left and right side panels respectively use the winding and Lagrangian based estimators for the correlation length, and the equation of state and field based estimators for the velocity; see the main text for detailed definitions of all the estimators. Low expansion rates are at the top of the panels while high expansion rates are at the bottom of the panels. All runs are of box size $512^3$, with constant comoving width, $s=0$ and $\alpha=2.0$. \label{fig2}}
\end{figure*}

There are then two possible estimators for the correlation length $\xi$. Since $\xi = \sqrt{\mathcal{V} / l}$ (with $\mathcal{V}$ and $l$ respectively being the box volume and the total length of string it contains) we need only find the total length of string in the box. The first estimator makes use of the fact that the Lagrangian density should vanish away from the string, while being negatively valued at the string itself \cite{Bevis:2006mj}; this leads to the definition
\begin{equation}
\xi_\mathcal{L} = \sqrt{ \frac{-\mu V}{\sum_x \mathcal{L}_x} }\,,
\end{equation}
which we will henceforth refer to as the Lagrangian-based correlation length estimator. The second estimator requires computing a gauge invariant winding \cite{Kajantie:1998bg} at each plaquette
\begin{equation}
W_{ij} = \frac{1}{2\pi} (Y_{i,x} + Y_{j,x+i} - Y_{i,x+j} -  Y_{j,x})\,,
\end{equation}
where $Y_i$ is given by
\begin{equation}
Y_i = [(\phi^x)_{arg} -(\phi^{x+k_i})_{arg} +A_{i,x}  ]_\pi - A_{i,x}\,.
\end{equation}
If around a given plaquette we have $W_{ij} \neq 0$, then a piece of string with length $\Delta x$ is present, so in order to obtain the total string length one only needs to sum $W_{ij}$ throughout the lattice
\begin{equation}
\xi_W = \sqrt{\frac{\mathcal{V}}{\sum_{ij,x} W_{ij,x}}}\,,
\end{equation}
which we will henceforth refer to as the winding-based correlation length estimator. Note that the obtained length is multiplied by a factor of $\pi/6$. This is necessary given that there is an overestimation of the string length in a cubic mesh---this is known as the Manhattan effect, see \cite{Scherrer:1997sq}.

For the $<v^2>$ estimators there are also two options. The first one comes from \cite{Hindmarsh:2008dw, Hindmarsh:2017qff} and is based on the fact that for the conjugate scalar field momentum, $\Pi$, the configuration for a moving string can be given Lorentz boosts of the static field configuration. A detailed derivation can be found in \cite{Hindmarsh:2017qff}. In our case we simply quote the estimator itself,
\begin{equation}
<v^2>_{\phi} = \frac{2R}{1+R}\,,
\end{equation}
where $R$ is given by
\begin{equation}
R = \frac{\sum_x |\Pi|^2 \mathcal{W}}{\sum_{x,i} |D^+_{x,i} \phi|^2 \mathcal{W} }
\end{equation}
and $\mathcal{W}$ is a weight function, meant to merely localize the estimators around strings; we will refer to this as the field-based velocity estimator. The second possibility is to use the equation of state estimator of \cite{Hindmarsh:2017qff}, in which the volume averages of the pressure and the density (each weighted by some weight function $\mathcal{W}$) then yield
\begin{equation}
<v^2>_{\omega} = \frac{1}{2} \bigg( 1+3\frac{\sum_x p_x \mathcal{W}_x}{\sum_x \rho_x  \mathcal{W}_x} \bigg)\,;
\end{equation}
we will refer to this as the equation of state based velocity estimator. As for the weight functions, we have chosen to use the Lagrangian, as was done in \cite{Hindmarsh:2008dw, Hindmarsh:2017qff}. We have also previously used this choice when validating our code, by comparing the code results with those in the previous literature, as described in \cite{Correia:2018gew}.

Our simulations were executed using a recently developed Graphics Processing Unit accelerated application on appropriate hardware: specifically two NVIDIA 1080Ti's with a recent multiGPU patch, and one NVIDIA Quadro P5000 in single GPU mode. The performance of the single GPU version is discussed in \cite{Correia:2018gew}, while the performance and scalability of a multi-GPU version will be discussed in a future publication. We take twelve initial conditions with gauge field and conjugate fields set to zero, with the scalar field having random phase and unit magnitude. These are then used to seed twelve runs at each expansion rate, in a range of 43 expansion rates $m$ in the range $[0.5,0.95]$. One reason to simulate faster expansion rates than radiation ($m=1/2$) and matter ($m=2/3$) is that such a properly calibrated model should be able to describe the onset of the recent acceleration of the universe---in the case of domain walls, it has been shown that a VOS model calibrated with constant expansion rates $m$ does accurately describe the radiation-to-matter transition \cite{Rybak1}. We emphasize that in our simulations the same set of twelve different initial conditions is used for all 43 different expansion rates, so indeed the only difference between each set of 12 runs is the expansion rate. 

A first production run, with the winding-based correlation length estimator and the equation of state based velocity estimator, is therefore composed of 516 runs. A second production run with the Lagrangian based correlation length estimator and the scalar field based velocity estimator was also done, bringing the total run count to 1032. With the hardware resources mentioned in the previous paragraph, each production run of 516 simulations was completed in about 8.6 hours of wall clock time. A summary of the results of these simulations is depicted in Fig. \ref{fig2}.

For the purpose of calibrating the VOS model one needs to first ascertain the constancy of $\dot{\xi}$ and $<v^2>$. A numerical technicality is the fact that in the simulations the expected scaling law is not of the form $\xi \propto \eta$  but instead $\xi \propto (\eta - \eta_0)$, with $\eta_0$ being a numerical offset. This offset merely depends on the initial conditions chosen for the simulation box, with the de-facto quantity of interest being the slope of $\xi$ as $\dot{\xi}$. As such we will take the asymptotic value of
\begin{equation}
\epsilon = \frac{\xi}{(\eta - \eta_0)(1-m)}\,,
\end{equation}
as the quantity of interest when calibrating the VOS model. (Recall that $m$ is the expansion rate in physical time, that is $a\propto t^m$.) To do so we compute an average offset for each expansion rate and use it to compute the mean asymptote and its uncertainties. For our choice of initial conditions we find that this is in the range from 37 to 48, thus very mildly dependent on the expansion rate. Note that we could modify the initial conditions such that $\eta_0$ is driven to zero but this would have to be done for each run, at each expansion rate, so it is not a useful strategy. In any case we have verified that this would lead us to the same scaling values. Specifically, we have verified this in the radiation era using the initial conditions of one our runs but with a period of damping evolution before the true radiation era evolution. (Some trial and error was involved in selecting the amount of time steps of damped evolution to drive the initial offset to zero.) In summary we expect the following scaling laws
\begin{align}
\xi \propto (\eta - \eta_0)^\mu && v^2 \propto \eta^\nu\,,
\end{align}
with the expected values of $\mu=1$ and $\nu=0$ if the network has reached scaling.

\begin{figure*}
\includegraphics[width=1.0\columnwidth]{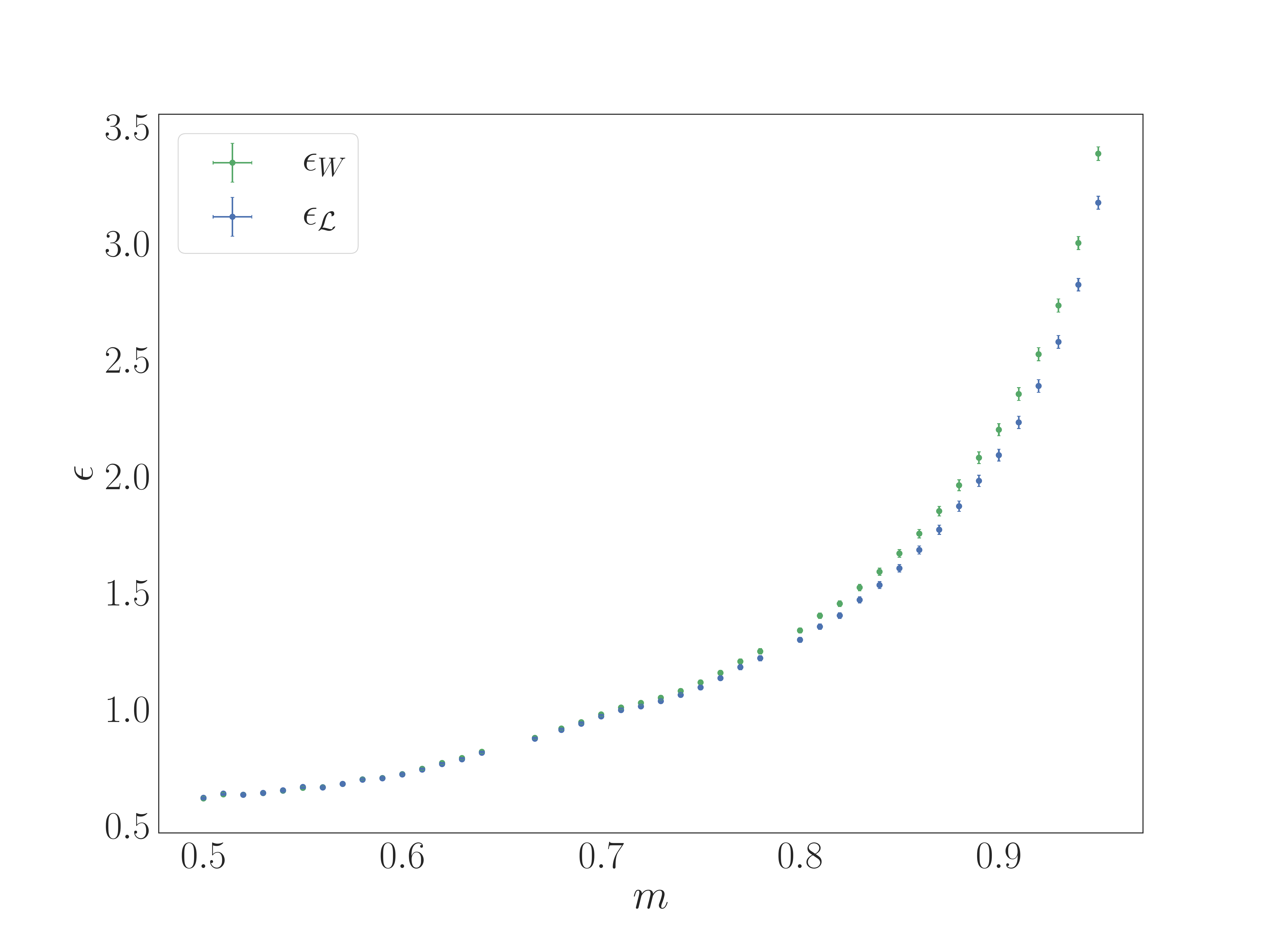}
\includegraphics[width=1.0\columnwidth]{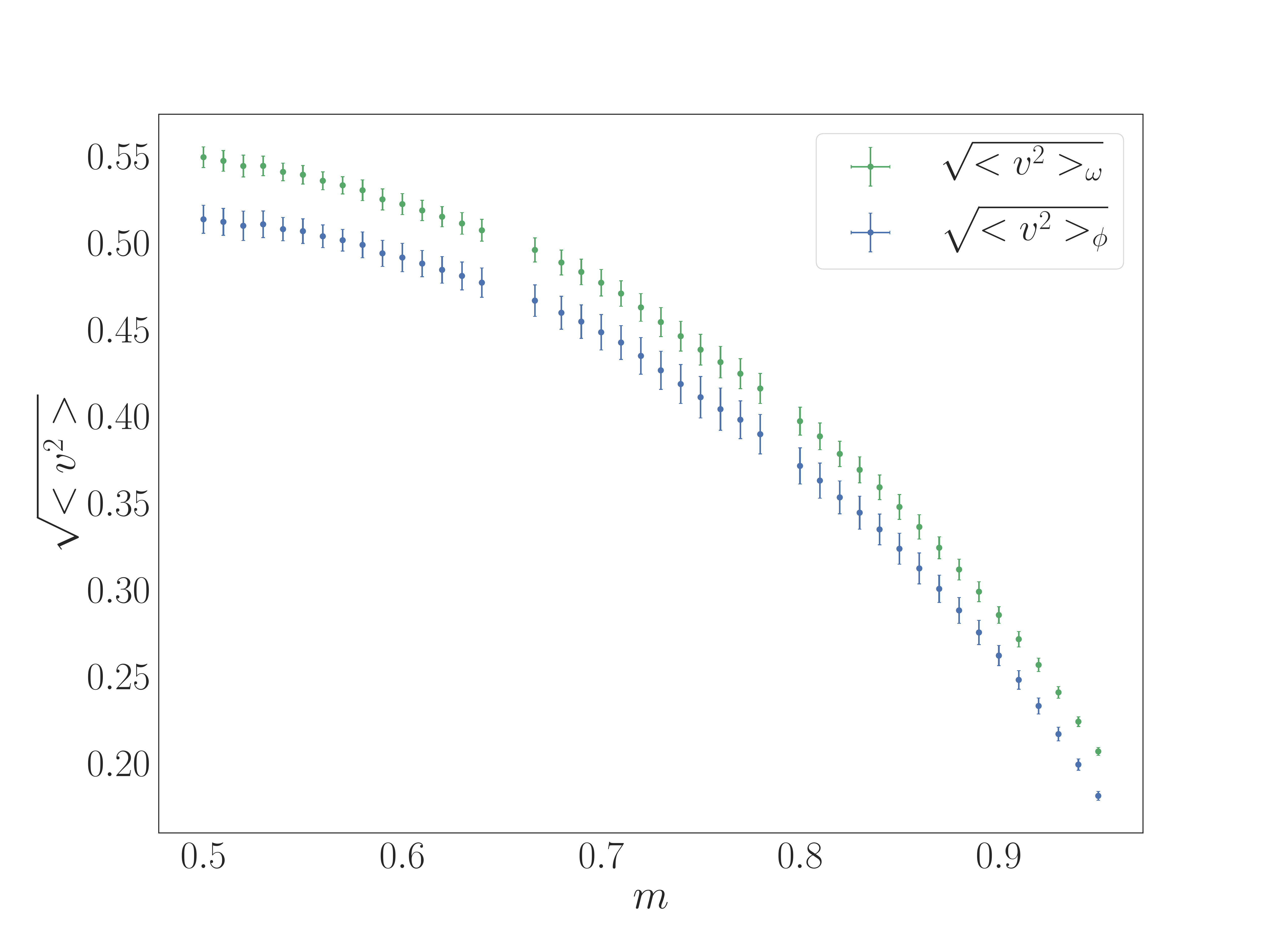}
\includegraphics[width=1.0\columnwidth]{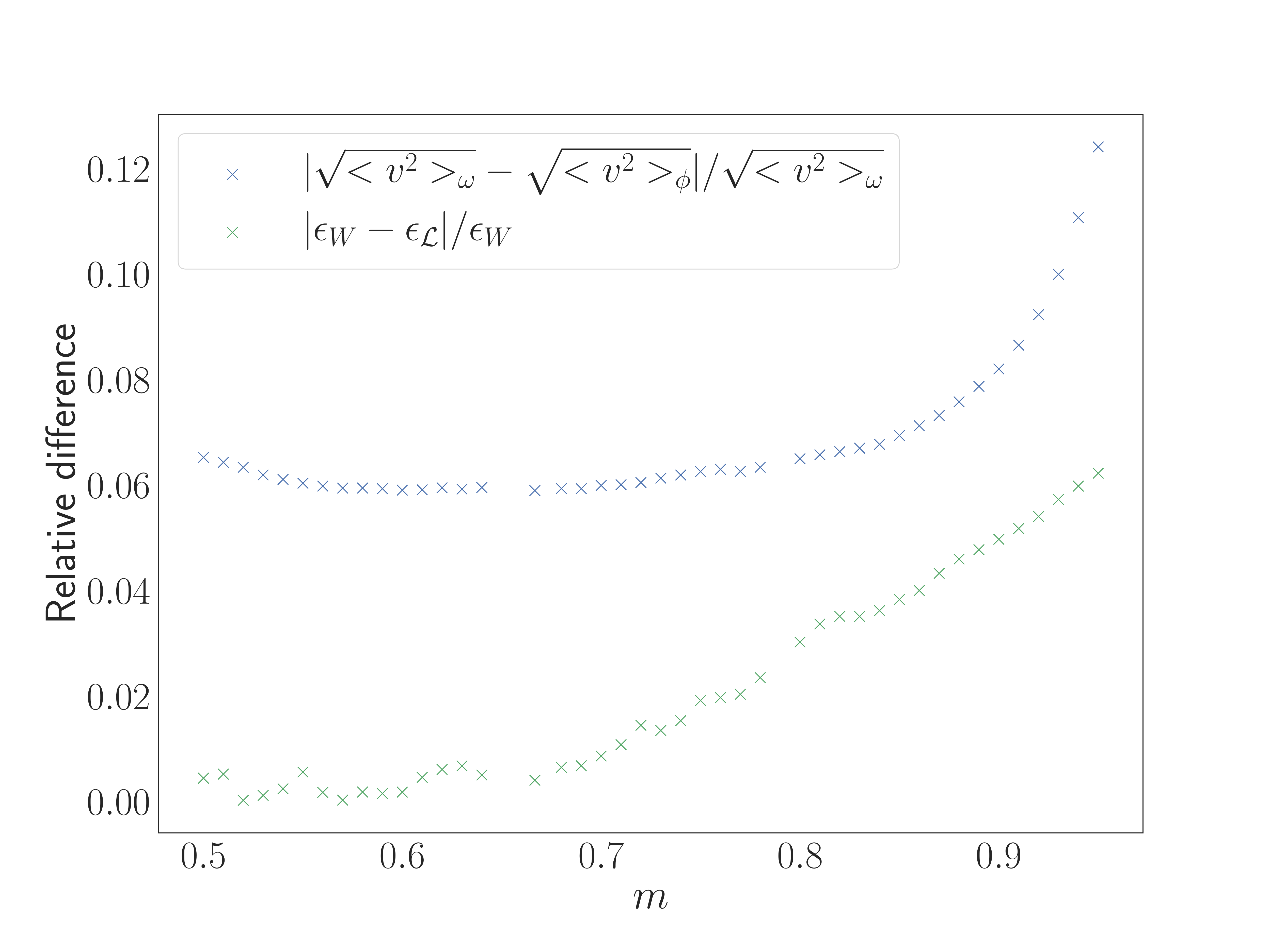}
\caption{Asymptotic values of $\epsilon$(top left panel) and root mean squared velocity $<v^2>$ (top right panel) for the two pairs of estimators used in the our production runs. The bottom panel shows the relative difference between the pairs of estimators, showing that the difference between the obtained velocities is in the range $6\%$-$-12\%$ while for the correlation length estimators it is at most of $6\%$.\label{fig3}}
\end{figure*}

The values obtained for these two exponents are listed, together with the related scaling parameters, in Table \ref{tab1}, and these scaling parameters are also depicted in Fig. \ref{fig3}. This information was used to select a timestep range to be used for VOS calibration, the requirement being that the scaling assumption holds to a sufficiently good approximation; see \cite{Rybak1,Rybak2} for a detailed discussion for the case of domain walls. Specifically, in the present work we have chosen to use timesteps in the time range of $\eta \in [80,128]$, corresponding to the final third of each simulation.

For this work, we will not use the field-based velocity estimator. The reason for this is that this estimator seems to systematically underestimate the velocity, which can also be seen in \cite{Hindmarsh:2017qff} when comparing to an oscillating string in flat space. The underestimation manifests itself in all expansion rates as seen in Fig. \ref{fig3} and Table \ref{tab1}, being more significant at large expansion rates; overall it ranges between six and twelve percent. The correlation length estimators seem to be in agreement at low expansion rates, when comparing $\epsilon$; the slight disagreement at larger expansion rates (at the level of a few percent) will lead to small differences in calibration between the two sets, as discussed in the following section.

For the scaling exponent $\mu$ all simulations are fully consistent with scaling (to three significant digits the values inferred are either 0.999 or 1.000 in all cases). For the corresponding exponent for the velocities, $\nu$, the values differ from the expected value of zero by typically a few percent (and maximally about ten percent). However, given the difficulties in numerically measuring the velocities (cf. the biases of the estimators discussed in the previous paragraph) we believe that this is not particularly significant. We thus operationally conclude that our networks have, for our purposes, reached the scaling regime in the range $\eta \in [80,128]$, and these data can therefore be used to calibrate the VOS model.

\begin{table*}
\begin{tabular}{| c | c c c c | c c c c|}
\hline
$m$ & $\mu_W$ &  $\nu_\omega$ &$\xi_W/(\eta-\eta_0)$ & $\sqrt{<v^2>_\omega}$ & $\mu_\mathcal{L}$ & $\nu_\phi$ &$\xi_\mathcal{L}/(\eta-\eta_0)$ & $\sqrt{<v^2>_\phi}$ \\
\hline
0.50   & 0.999$\pm$0.005 & 0.024$\pm$0.004 & 0.307$\pm$0.004 & 0.549$\pm$0.006&	1.000$\pm$0.005 & 0.047$\pm$0.007 & 0.309$\pm$0.004 & 0.513$\pm$0.008\\
0.51   & 1.000$\pm$0.005 & 0.003$\pm$0.005 & 0.310$\pm$0.004 & 0.547$\pm$0.006&	1.000$\pm$0.005 & 0.014$\pm$0.006 & 0.311$\pm$0.004 & 0.512$\pm$0.008\\
0.52   & 0.999$\pm$0.005 & 0.003$\pm$0.005 & 0.303$\pm$0.004 & 0.544$\pm$0.006&	0.999$\pm$0.005 & 0.023$\pm$0.007 & 0.303$\pm$0.004 & 0.510$\pm$0.009\\
0.53   & 0.999$\pm$0.005 & 0.008$\pm$0.004 & 0.300$\pm$0.004 & 0.544$\pm$0.006&	0.999$\pm$0.005 & 0.027$\pm$0.006 & 0.300$\pm$0.004 & 0.510$\pm$0.008\\
0.54   & 0.999$\pm$0.004 & 0.004$\pm$0.004 & 0.298$\pm$0.003 & 0.541$\pm$0.005&	0.999$\pm$0.004 & 0.019$\pm$0.006 & 0.299$\pm$0.003 & 0.508$\pm$0.007\\
0.55   & 0.999$\pm$0.004 & 0.017$\pm$0.004 & 0.297$\pm$0.003 & 0.539$\pm$0.005&	1.000$\pm$0.004 & 0.034$\pm$0.006 & 0.299$\pm$0.003 & 0.506$\pm$0.007\\
0.56   & 0.999$\pm$0.003 & 0.009$\pm$0.004 & 0.292$\pm$0.002 & 0.536$\pm$0.005&	0.999$\pm$0.003 & 0.024$\pm$0.005 & 0.291$\pm$0.002 & 0.504$\pm$0.007\\
0.57   & 0.999$\pm$0.003 & 0.023$\pm$0.004 & 0.291$\pm$0.002 & 0.533$\pm$0.005&	0.999$\pm$0.003 & 0.043$\pm$0.005 & 0.291$\pm$0.002 & 0.501$\pm$0.006\\
0.58   & 0.999$\pm$0.003 & 0.036$\pm$0.005 & 0.292$\pm$0.002 & 0.530$\pm$0.006&	0.999$\pm$0.003 & 0.057$\pm$0.006 & 0.292$\pm$0.002 & 0.499$\pm$0.007\\
0.59   & 0.999$\pm$0.003 & 0.033$\pm$0.005 & 0.288$\pm$0.003 & 0.525$\pm$0.006&	0.999$\pm$0.003 & 0.054$\pm$0.006 & 0.287$\pm$0.003 & 0.494$\pm$0.008\\
0.60   & 0.999$\pm$0.003 & 0.027$\pm$0.005 & 0.288$\pm$0.002 & 0.522$\pm$0.006&	0.999$\pm$0.003 & 0.045$\pm$0.007 & 0.287$\pm$0.003 & 0.491$\pm$0.008\\
0.61   & 0.999$\pm$0.003 & 0.029$\pm$0.005 & 0.289$\pm$0.002 & 0.518$\pm$0.006&	0.999$\pm$0.003 & 0.046$\pm$0.006 & 0.288$\pm$0.003 & 0.488$\pm$0.008\\
0.62   & 0.999$\pm$0.003 & 0.043$\pm$0.005 & 0.291$\pm$0.002 & 0.515$\pm$0.006&	0.999$\pm$0.003 & 0.060$\pm$0.007 & 0.290$\pm$0.002 & 0.484$\pm$0.008\\
0.63   & 0.999$\pm$0.003 & 0.051$\pm$0.005 & 0.292$\pm$0.003 & 0.511$\pm$0.006&	0.999$\pm$0.003 & 0.066$\pm$0.007 & 0.290$\pm$0.003 & 0.481$\pm$0.008\\
0.64   & 0.999$\pm$0.003 & 0.054$\pm$0.005 & 0.293$\pm$0.003 & 0.507$\pm$0.006&	0.999$\pm$0.003 & 0.073$\pm$0.007 & 0.292$\pm$0.003 & 0.477$\pm$0.008\\
0.6(6) & 1.000$\pm$0.003 & 0.073$\pm$0.006 & 0.292$\pm$0.002 & 0.496$\pm$0.007&	1.000$\pm$0.003 & 0.091$\pm$0.008 & 0.290$\pm$0.002 & 0.466$\pm$0.009\\
0.68   & 0.999$\pm$0.003 & 0.070$\pm$0.006 & 0.293$\pm$0.002 & 0.488$\pm$0.007&	1.000$\pm$0.003 & 0.089$\pm$0.009 & 0.291$\pm$0.002 & 0.459$\pm$0.010\\
0.69   & 0.999$\pm$0.003 & 0.080$\pm$0.006 & 0.292$\pm$0.003 & 0.483$\pm$0.007&	1.000$\pm$0.003 & 0.102$\pm$0.009 & 0.290$\pm$0.003 & 0.454$\pm$0.010\\
0.70   & 1.000$\pm$0.003 & 0.085$\pm$0.006 & 0.293$\pm$0.002 & 0.477$\pm$0.008&	1.000$\pm$0.003 & 0.107$\pm$0.010 & 0.290$\pm$0.002 & 0.448$\pm$0.010\\
0.71   & 1.000$\pm$0.003 & 0.084$\pm$0.006 & 0.291$\pm$0.002 & 0.471$\pm$0.007&	1.000$\pm$0.003 & 0.105$\pm$0.009 & 0.288$\pm$0.002 & 0.442$\pm$0.010\\
0.72   & 0.999$\pm$0.003 & 0.081$\pm$0.007 & 0.287$\pm$0.002 & 0.463$\pm$0.008&	0.999$\pm$0.003 & 0.106$\pm$0.010 & 0.283$\pm$0.002 & 0.435$\pm$0.011\\
0.73   & 0.999$\pm$0.004 & 0.083$\pm$0.007 & 0.283$\pm$0.003 & 0.454$\pm$0.008&	0.999$\pm$0.003 & 0.106$\pm$0.011 & 0.279$\pm$0.003 & 0.426$\pm$0.011\\
0.74   & 0.999$\pm$0.004 & 0.074$\pm$0.008 & 0.280$\pm$0.003 & 0.446$\pm$0.009&	1.000$\pm$0.004 & 0.091$\pm$0.012 & 0.275$\pm$0.003 & 0.418$\pm$0.011\\
0.75   & 0.999$\pm$0.004 & 0.060$\pm$0.009 & 0.278$\pm$0.003 & 0.438$\pm$0.009&	0.999$\pm$0.003 & 0.071$\pm$0.014 & 0.273$\pm$0.002 & 0.411$\pm$0.012\\
0.76   & 0.999$\pm$0.004 & 0.062$\pm$0.009 & 0.277$\pm$0.003 & 0.431$\pm$0.009&	1.000$\pm$0.003 & 0.072$\pm$0.014 & 0.272$\pm$0.002 & 0.404$\pm$0.012\\
0.77   & 1.000$\pm$0.004 & 0.083$\pm$0.008 & 0.277$\pm$0.003 & 0.424$\pm$0.009&	1.000$\pm$0.004 & 0.094$\pm$0.012 & 0.271$\pm$0.003 & 0.398$\pm$0.011\\
0.78   & 1.000$\pm$0.004 & 0.084$\pm$0.009 & 0.274$\pm$0.003 & 0.416$\pm$0.009&	1.000$\pm$0.003 & 0.095$\pm$0.013 & 0.268$\pm$0.002 & 0.389$\pm$0.011\\
0.80   & 1.000$\pm$0.003 & 0.075$\pm$0.008 & 0.267$\pm$0.002 & 0.397$\pm$0.008&	1.000$\pm$0.003 & 0.083$\pm$0.012 & 0.259$\pm$0.002 & 0.371$\pm$0.010\\
0.81   & 1.000$\pm$0.003 & 0.074$\pm$0.008 & 0.266$\pm$0.002 & 0.388$\pm$0.008&	1.000$\pm$0.003 & 0.079$\pm$0.012 & 0.257$\pm$0.002 & 0.363$\pm$0.010\\
0.82   & 1.000$\pm$0.003 & 0.078$\pm$0.008 & 0.261$\pm$0.002 & 0.378$\pm$0.007&	1.000$\pm$0.003 & 0.083$\pm$0.012 & 0.252$\pm$0.002 & 0.353$\pm$0.010\\
0.83   & 1.000$\pm$0.003 & 0.091$\pm$0.008 & 0.259$\pm$0.002 & 0.369$\pm$0.007&	1.000$\pm$0.004 & 0.098$\pm$0.012 & 0.249$\pm$0.002 & 0.344$\pm$0.009\\
0.84   & 1.000$\pm$0.004 & 0.101$\pm$0.008 & 0.254$\pm$0.002 & 0.359$\pm$0.007&	1.000$\pm$0.004 & 0.109$\pm$0.011 & 0.245$\pm$0.002 & 0.335$\pm$0.009\\
0.85   & 1.000$\pm$0.004 & 0.106$\pm$0.008 & 0.250$\pm$0.002 & 0.347$\pm$0.007&	1.000$\pm$0.004 & 0.116$\pm$0.012 & 0.240$\pm$0.002 & 0.323$\pm$0.009\\
0.86   & 1.000$\pm$0.004 & 0.102$\pm$0.008 & 0.245$\pm$0.003 & 0.336$\pm$0.007&	1.000$\pm$0.004 & 0.109$\pm$0.012 & 0.235$\pm$0.002 & 0.312$\pm$0.009\\
0.87   & 1.000$\pm$0.004 & 0.101$\pm$0.008 & 0.240$\pm$0.003 & 0.324$\pm$0.006&	1.000$\pm$0.004 & 0.108$\pm$0.011 & 0.230$\pm$0.003 & 0.300$\pm$0.008\\
0.88   & 1.000$\pm$0.005 & 0.095$\pm$0.008 & 0.235$\pm$0.003 & 0.311$\pm$0.006&	1.000$\pm$0.005 & 0.098$\pm$0.011 & 0.224$\pm$0.003 & 0.288$\pm$0.007\\
0.89   & 1.000$\pm$0.005 & 0.091$\pm$0.007 & 0.229$\pm$0.003 & 0.299$\pm$0.006&	1.000$\pm$0.005 & 0.090$\pm$0.010 & 0.218$\pm$0.003 & 0.275$\pm$0.007\\
0.90   & 1.000$\pm$0.004 & 0.092$\pm$0.006 & 0.220$\pm$0.003 & 0.285$\pm$0.005&	1.000$\pm$0.005 & 0.089$\pm$0.009 & 0.209$\pm$0.003 & 0.262$\pm$0.006\\
0.91   & 1.000$\pm$0.004 & 0.097$\pm$0.006 & 0.212$\pm$0.002 & 0.271$\pm$0.004&	1.000$\pm$0.004 & 0.093$\pm$0.009 & 0.201$\pm$0.002 & 0.248$\pm$0.005\\
0.92   & 1.000$\pm$0.004 & 0.106$\pm$0.005 & 0.202$\pm$0.002 & 0.256$\pm$0.004&	1.000$\pm$0.004 & 0.106$\pm$0.008 & 0.191$\pm$0.002 & 0.233$\pm$0.005\\
0.93   & 1.000$\pm$0.004 & 0.097$\pm$0.005 & 0.191$\pm$0.002 & 0.241$\pm$0.003&	1.000$\pm$0.004 & 0.097$\pm$0.007 & 0.180$\pm$0.002 & 0.217$\pm$0.004\\
0.94   & 0.999$\pm$0.003 & 0.077$\pm$0.005 & 0.180$\pm$0.002 & 0.224$\pm$0.003&	0.999$\pm$0.004 & 0.070$\pm$0.006 & 0.169$\pm$0.002 & 0.199$\pm$0.003\\
0.95   & 0.999$\pm$0.003 & 0.070$\pm$0.004 & 0.169$\pm$0.001 & 0.207$\pm$0.002&	0.999$\pm$0.003 & 0.053$\pm$0.006 & 0.159$\pm$0.001 & 0.181$\pm$0.003\\
\hline
\end{tabular}
\caption{Relevant quantities measured from the two sets of simulations, for each expansion rate $m$: specifically the scaling exponents, $\mu$ and $\nu$, together with the mean correlation length divided by conformal time (corrected by an offset), $\xi/(\tau-\tau_0)$, and the mean velocity squared $<v^2>$. The left side of the table uses the winding-based correlation length estimator and the equation of state based velocity estimator, while the right side of the table uses the Lagrangian-based correlation length estimator and the field-based velocity estimator. All quantities are the result of the average of 12 simulations with different initial conditions.\label{tab1}}
\end{table*}

\section{\label{evos}Extending the Velocity-dependent One-Scale Model}

The VOS model is the canonical analytic approach to treating cosmic string network evolution. For detailed derivations we refer the reader to the original works \cite{Martins:1996jp,MS2} and to a recent overview \cite{Book}. In what follows we will limit ourselves to introducing the evolution equations for the model's parameters, the average correlation length $L$ and the root-mean square velocity $v$ of a string network. Note that we temporarily retain the standard symbols $L$ and $v$ for these, both because they are standard in the literature and to distinguish them from the numerically measured ones, $\xi$ and $<v^2>$, to be described in what follows.

Written in terms of physical time $t$, the model's evolution equations are
\begin{equation}
2\frac{dL}{dt} = 2HL(1+v^2) + F(v)
\end{equation}
\begin{equation}
\frac{dv}{dt} = \bigg( 1-v^2 \bigg) \bigg( \frac{k(v)}{L} -2Hv \bigg)\,,
\end{equation}
where $H$ is the Hubble parameter, $k(v)$ is the momentum parameter (which encodes a phenomenological description of small-scale structures on the strings), and $F(v)=cv$ is the loop production term, with $c$ being known as the loop chopping efficiency. Note that in the original VOS model the only energy loss term is due to loop production. The momentum parameter is described in detail in \cite{MS2}; in general it has been inferred to have the form
\begin{equation} \label{eq:kvfull}
k(v) =  \frac{2\sqrt{2}}{\pi} (1-v^2)(1+2\sqrt{2}v^3)\frac{1-8v^6}{1+8v^6}\,.
\end{equation}
This is chosen to interpolate between the non-relativistic (friction-dominated) and relativistic limits, mostly by comparison to Goto-Nambu simulations. Note that the momentum parameter is maximal in the non-relativistic limit, and vanishes for a maximal velocity which in the above has been assumed to be $v^2=1/2$. In the relativistic case (which is the focus of our attention in the current work), the simpler form 
\begin{equation} \label{eq:kv}
k(v) =  \frac{2\sqrt{2}}{\pi}\frac{1-8v^6}{1+8v^6}
\end{equation}
is usually adequate, and has often been used in previous literature.

\begin{figure*}
\includegraphics[width=1.0\columnwidth]{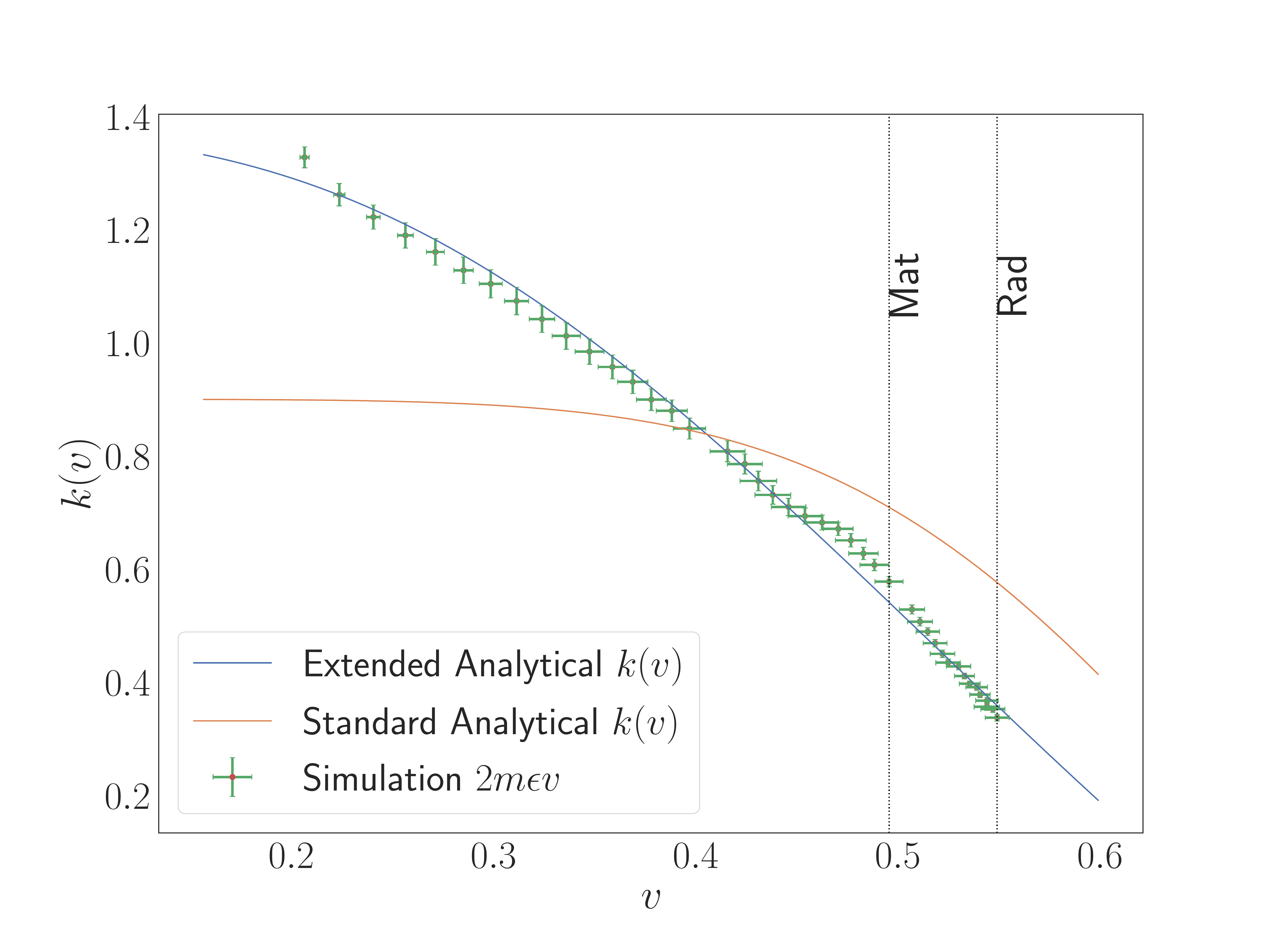}
\includegraphics[width=1.0\columnwidth]{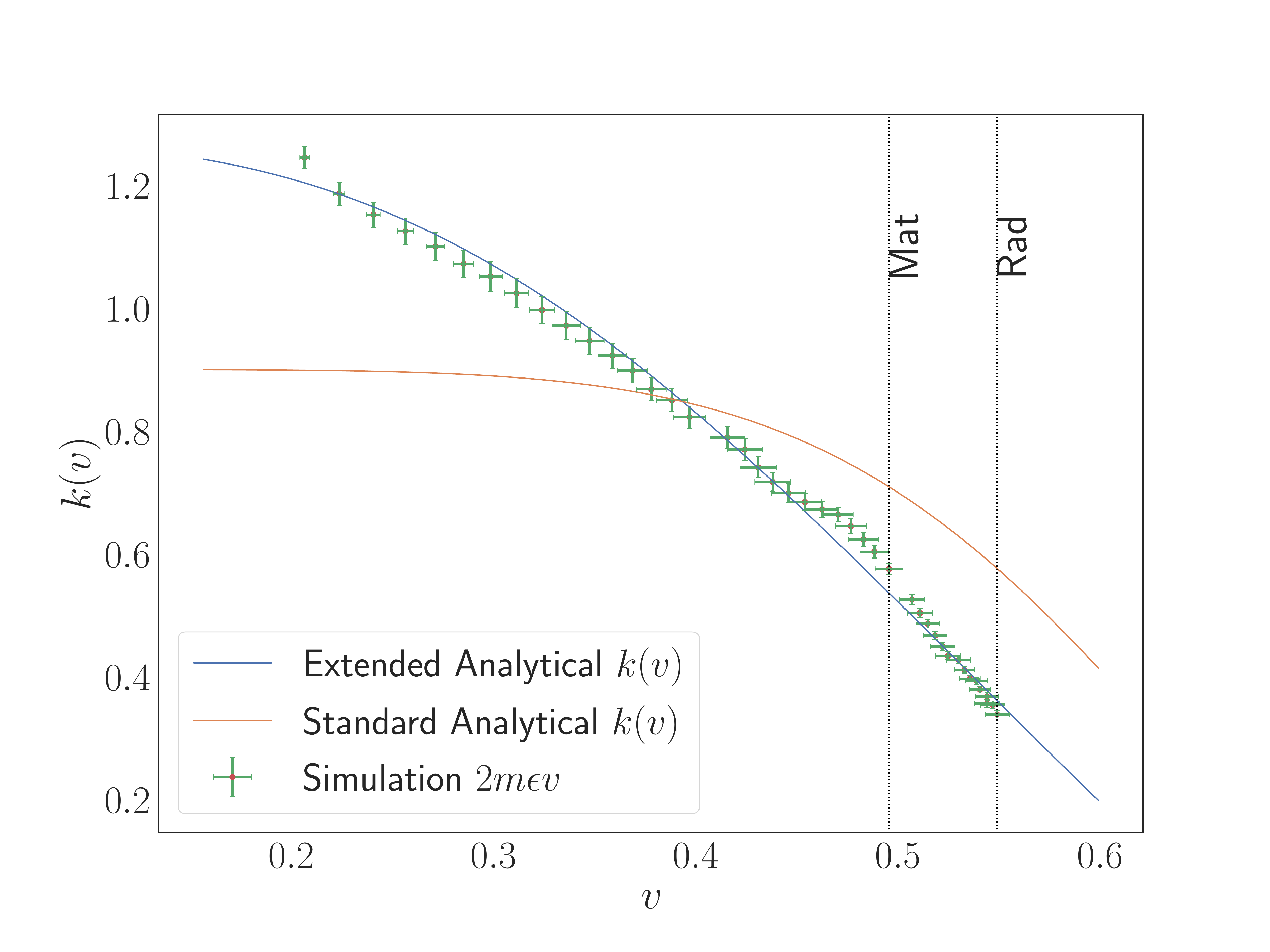}
\includegraphics[width=1.0\columnwidth]{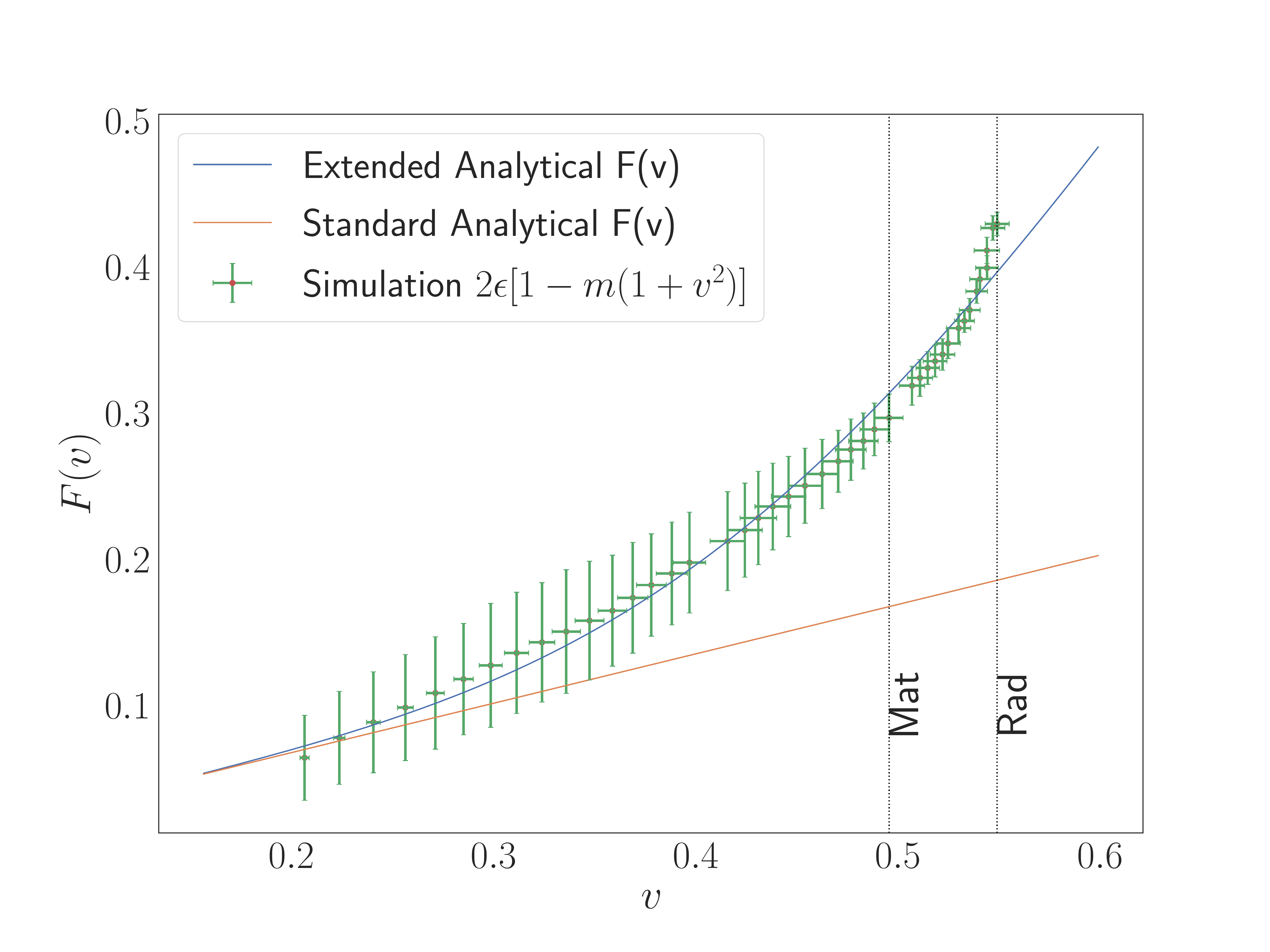}
\includegraphics[width=1.0\columnwidth]{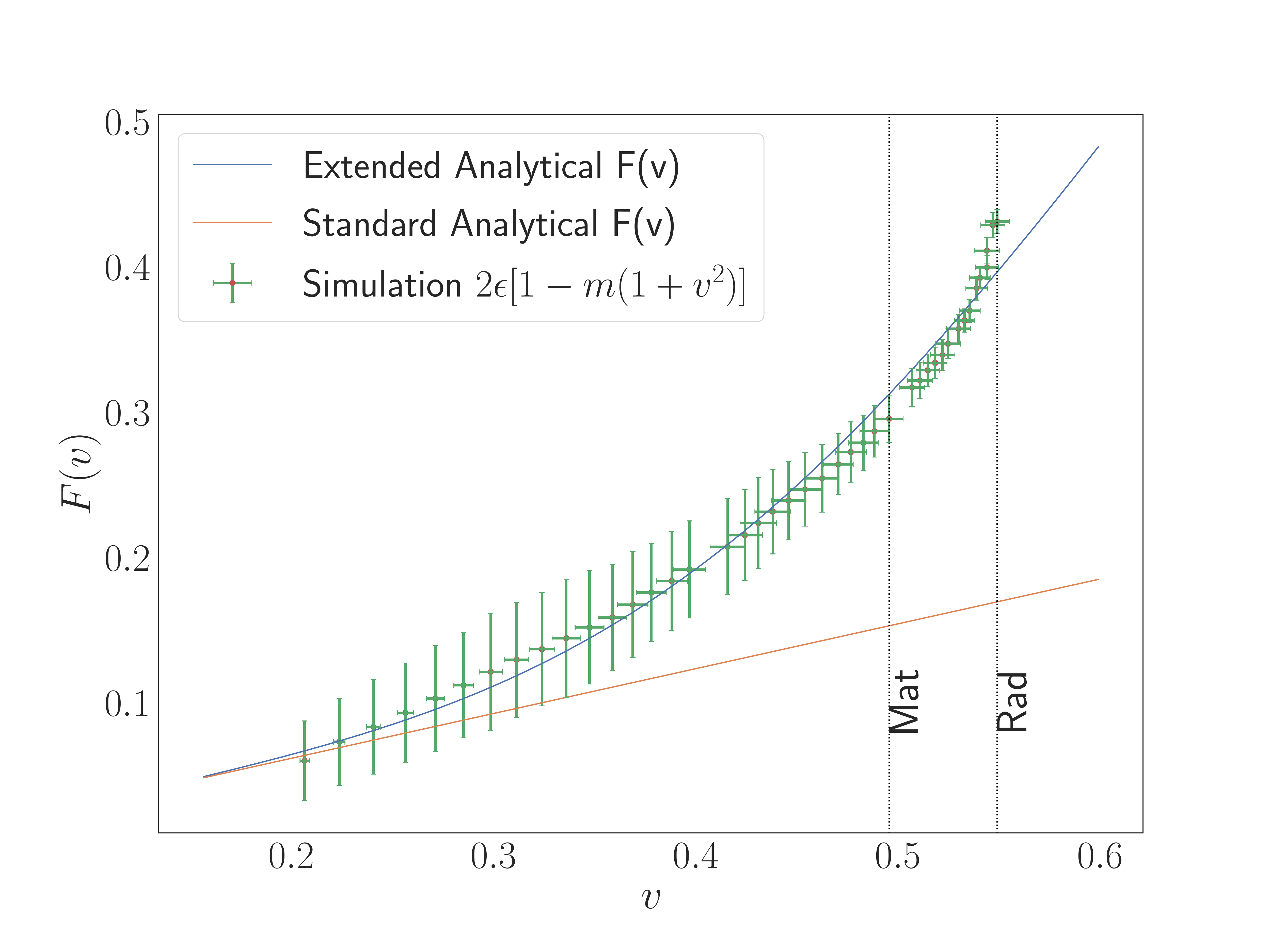}
\caption{Comparisons between the analytic VOS model predictions (solid lines) and the simulation outputs (data points) for both the momentum parameter $k(v)$ (top panels) and a generalized energy loss function $F(v)$ (bottom panels). Left side and right side panels correspond to the winding-based and Lagrangian-based correlation length estimators discussed in the text. In each case we show the simulation diagnostics used as input for the inverted VOS expressions. We show for comparison both the previous and extended versions of $k(v)$ and $F(v)$ (depicted in red and blue lines, and given respectively by Eqs. \ref{eq:vos1}--\ref{eq:vos2} and Eqs. \ref{eq:vos3}--\ref{eq:vos4}) in order to emphasize that the previous one provides a poor fit while the extended one provides a very good one. To facilitate comparisons with previous works the radiation and matter era values are explicitly indicated.\label{fig4}}
\end{figure*}

Since we evolve our simulations in conformal time, $\eta$, and thus measure the conformal correlation length, $\xi$, it is now convenient to rewrite the VOS evolution equations as
\begin{equation}
\frac{d\xi}{d\eta} = \frac{m}{(1-m)\eta} v^2 +\frac{cv}{2}
\end{equation}
\begin{equation}
\frac{dv}{d\eta} = (1-v^2) \bigg[ \frac{k(v)}{\xi}  - \frac{2m v}{(1-m)\eta} \bigg]\,.
\end{equation}
Moreover, since we will calibrate this model in the linear scaling regime, we can further rewrite the above equations using asymptotic quantities, $v_0$ and $\epsilon$, as
\begin{equation}\label{eq:vos1}
cv_0 = 2 \epsilon [1-m (1-v_0^2)]
\end{equation}
\begin{equation}\label{eq:vos2}
k(v) = 2m \epsilon v_0\,.
\end{equation}
At this point, one is ready to compare the analytic assumptions for the momentum parameter and the energy loss term directly with simulation output. This comparison can be seen in Fig. \ref{fig4}---refer to the solid orange lines therein. As has been previously noticed for the case of domain walls \cite{Rybak1}, the standard VOS model (which previously had been calibrated using only radiation and matter era simulations, in addition to Minkowski spacetime ones) does not provide a good fit to the extended range of expansion rates.

We thus propose to extend this model by taking inspiration from the recent extension of the VOS model for domain walls \cite{Rybak1,Rybak2}, specifically by considering a more general momentum parameter, and by further allowing for energy losses due to scalar or gauge radiation. The generalized momentum parameter is defined to be
 \begin{equation}\label{eq:vos3}
k(v) =  k_0\frac{1-(qv^2)^{\beta}}{1+(qv^2)^{\beta}}\,,
\end{equation}
where $\beta$, $q$ and $k_0$ are free parameters to be determined from the simulations. Note that appropriate choices of the aforementioned parameters reduce this $k(v)$ to the ansatz of Eq. \ref{eq:kv}. It is also worthy of note that for a non-wiggly string $k_0$ cannot be more than unity, but such a restriction does not hold for wiggly strings. The energy loss term will be modified to include the scalar radiation term
\begin{equation}\label{eq:vos4}
F(v) = \frac{cv}{2} + \frac{d[k_0-k(v)]^r}{2}
\end{equation}
where $d$ and $r$ are additional free parameters. Note that $c$ and $d$ quantify the relative contributions of loop production and scalar and gauge radiation. The form of the new term stems from the expectation that uniformly moving defects do not radiate---only perturbations of the defect surface will provide such radiation. One expects that radiation will be proportionally more important for slow expansion rates (corresponding to larger defect velocities), while the loop chopping term will be proportionally more important for faster expansions.

This extension of the VOS model allows us to phenomenologically test which energy loss mechanism is the dominant one and also to test if the ansatz for the relativistic momentum parameter is a reasonable assumption for a realistic network of strings. (In future work we will extend the present analysis and test the non-relativistic case.) Moreover, a comparative analysis can also be done with the analogous VOS model for domain walls. Clearly the model now has a significant number---six---of phenomenological parameters, but this is not a problem \textit{per se}: as we will show in the following section, having simulations with a large number of different expansion rates allows us to numerically measure each of these parameters with a very good level of statistical significance.

\section{\label{calib}Calibrating the extended VOS model}

Our extended VOS model can now be calibrated using the previously described GPU-based simulations data. As was done for the domain walls case \cite{Rybak1,Rybak2}, this can be done by following a standard bootstrap procedure. We will separately consider the winding and Lagrangian estimators for the correlation length, in both cases using the equation of state estimator, for the previously mentioned reasons. The calibrated model parameters and their corresponding uncertainties are summarized in Table \ref{tab2}, in which for comparison we also list the analogous results for the domain walls VOS model, both for a range of expansion rates comparable to the one in the current work and thus containing only relativistic networks including radiation and matter era ones (coming from \cite{Rybak1}) and for an even wider range of expansion rates, also including ultra-relativistic and non-relativistic networks (coming from \cite{Rybak2}).

\begin{table*}
\begin{tabular}{| c || c | c || c | c |}
\hline
Parameter & Cosmic strings (Winding) & Cosmic strings (Lagrangian) & Domain walls (Relativistic) & Domain walls (All) \\
\hline
Reference & {\bf This work} &  {\bf This work} & \cite{Rybak1} & \cite{Rybak2} \\
\hline
$m$ range & [0.50-0.95] & [0.50-0.95] & [0.50-0.90] & [0.20-0.9998] \\
\hline
$k_0$ & $1.37\pm0.07$ & $1.27\pm0.09$ & $1.72\pm0.03$ & $1.77\pm0.03$ \\
$q$ & $2.30\pm0.04$ & $2.27\pm0.05$ & $4.10\pm0.17$ & $3.35\pm0.32$ \\
\hline
$\beta$ & $1.46\pm0.07$ & $1.54\pm0.09$ & $1.65\pm0.12$ & $1.08\pm0.07$ \\
$r$ & $1.85\pm0.11$ & $1.66\pm0.10$ & $1.30\pm0.06$ & $1.42\pm0.04$ \\
\hline
$d$ & $0.21\pm0.01$ & $0.26\pm0.01$ & $0.29\pm0.01$ & $0.26\pm0.02$ \\
$c$ & $0.34\pm0.02$ & $0.31\pm0.02$ & $0.00\pm0.03$ & $0.00\pm0.08$ \\
\hline
\end{tabular}
\caption{Calibrated parameters for the cosmic strings VOS model, obtained from the two sets of GPU-based simulations in this work and corresponding to the winding-based and Lagrangian-based correlation length estimators described in the text. For comparison we show the analogous parameters for the domain walls VOS model (obtained in previous work), both for a range of expansion rates comparable to the one in the present work and for a wider range of expansion rates.\label{tab2}}
\end{table*}

A first comment is that the winding and Lagrangian based correlation length estimators lead to VOS model parameters that are discernably different but nevertheless quite compatible with each other given the inferred uncertainties on the parameters. The only small difference occurs for the parameter $d$, but still this difference is only at the level of two standard deviations, and thus not statistically significant.

\begin{figure*}
\includegraphics[width=1.0\columnwidth]{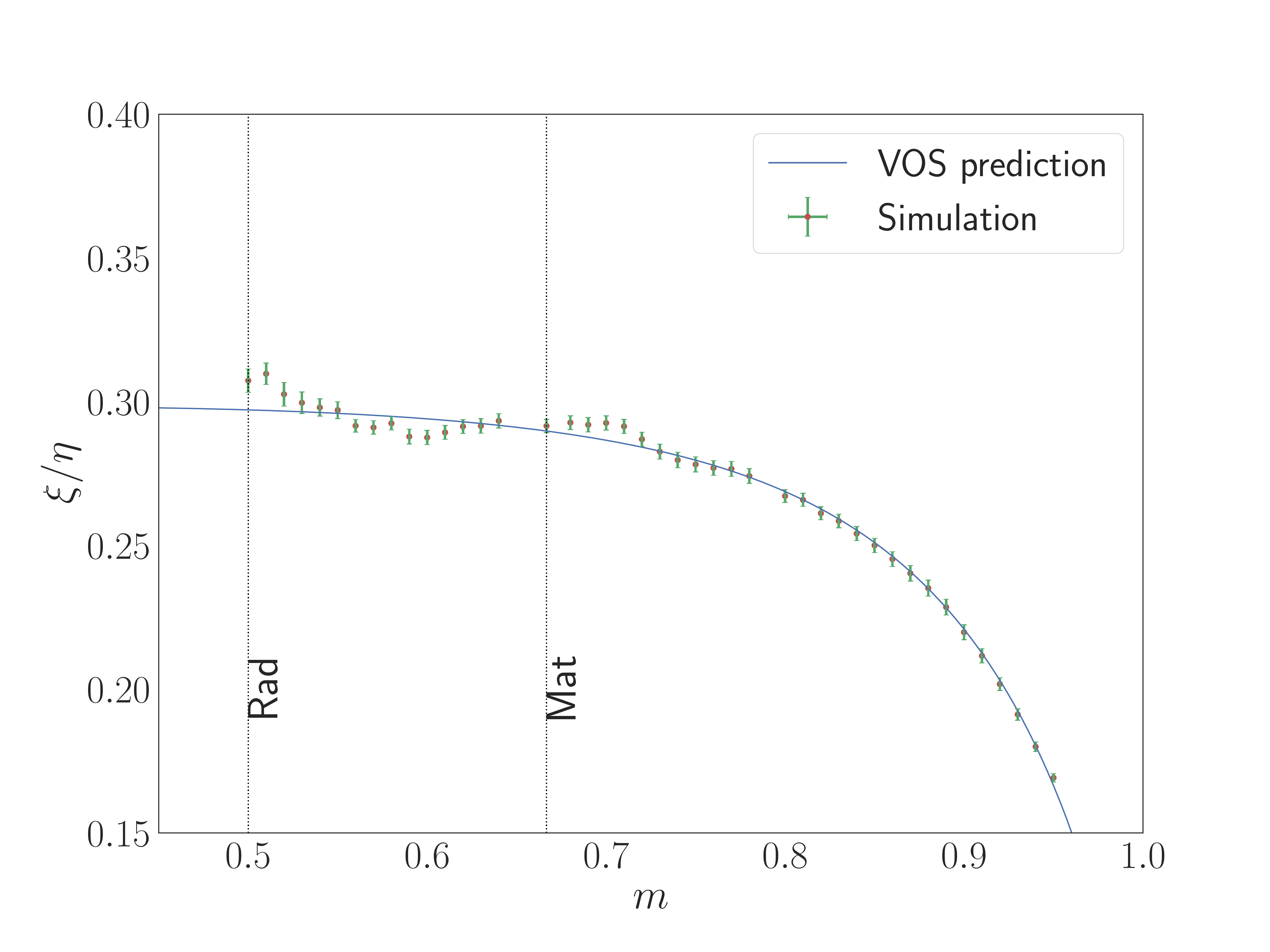}
\includegraphics[width=1.0\columnwidth]{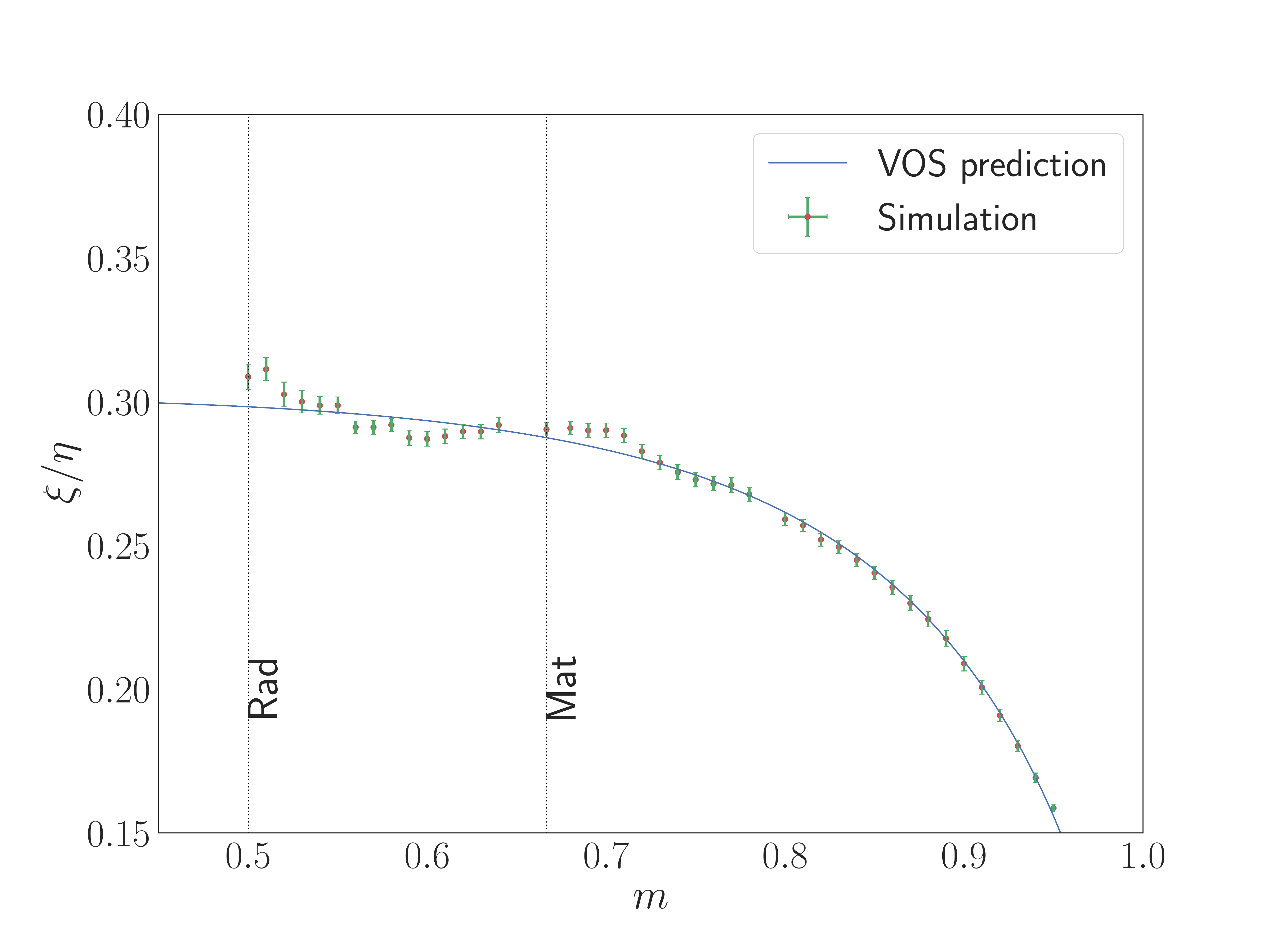}
\includegraphics[width=1.0\columnwidth]{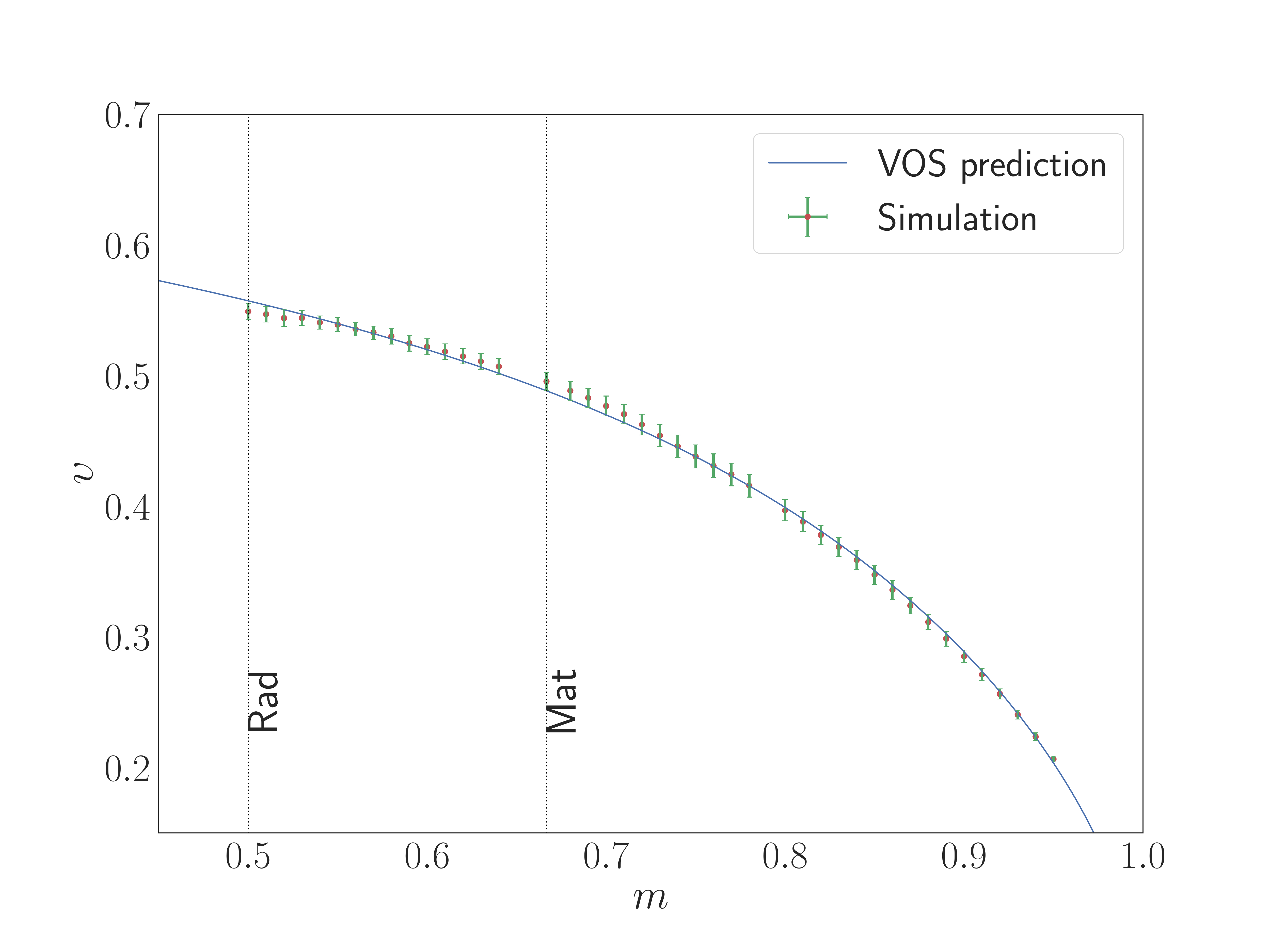}
\includegraphics[width=1.0\columnwidth]{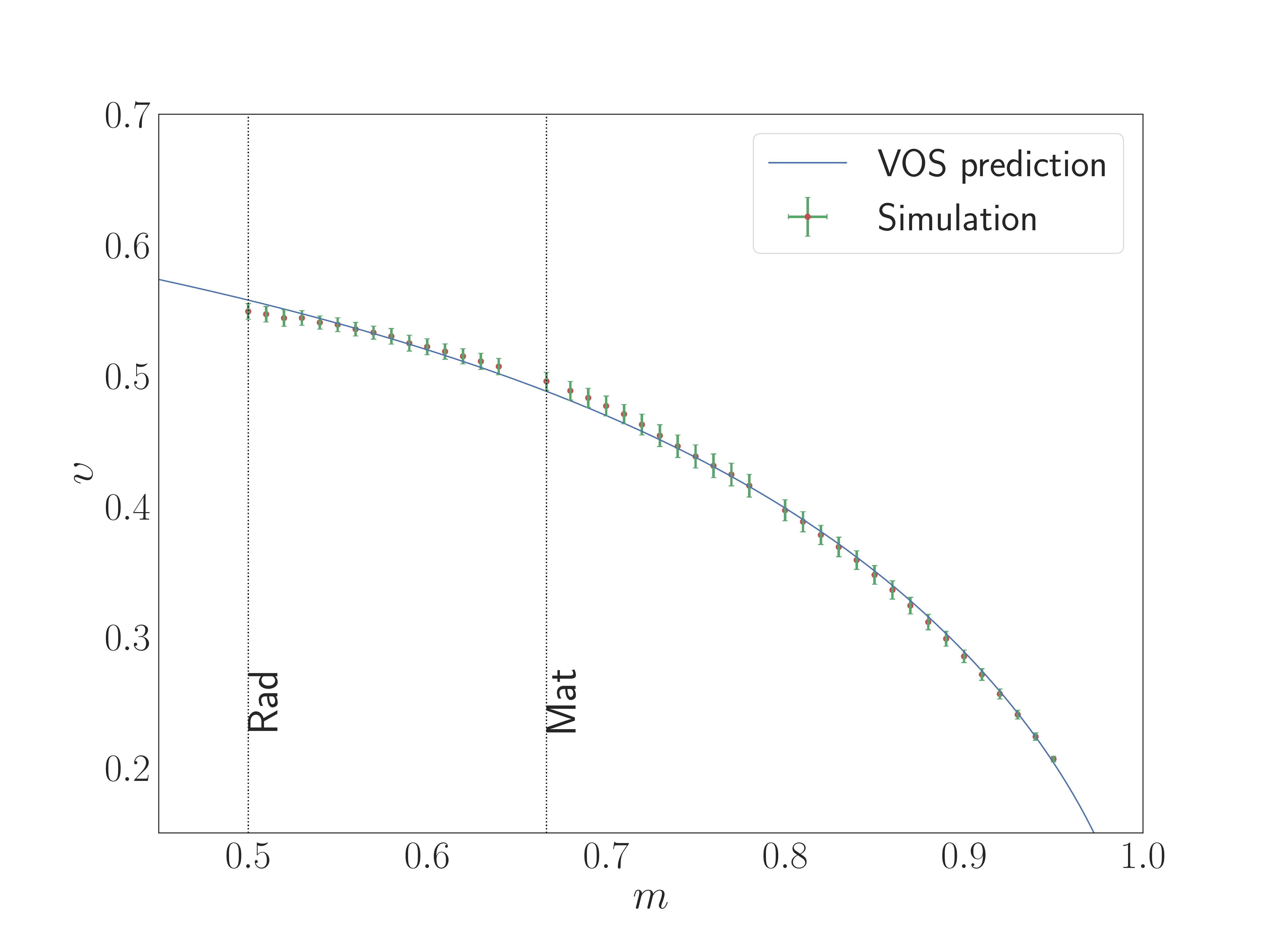}
\caption{Comparison between simulation outputs and the calibrated extended VOS model prediction for the rate of change of $\xi$, (specifically $\xi/\eta$, top panels) and the root mean square velocity (bottom panel). Left side and right side panels correspond to the two different choices of correlation length estimators, winding-based and Lagrangian-based, described in the text. To facilitate comparisons with previous works the radiation and matter era values are explicitly indicated.\label{fig5}}
\end{figure*}

A second noteworthy feature of this set of parameters is that the preferred value of $\beta$ disagrees with the one corresponding to the analytic ansatz of Eq. \ref{eq:kv}, which would be $\beta=3$; instead, a value around $\beta=1.5$ is preferred. This is certainly part of the reason why the previous version of the VOS model does not accurately reproduce the velocity dependencies of the relevant dynamical quantities (cf. the solid orange lines in Fig. \ref{fig4}), though note that an additional reason is that there are some degeneracies between the model parameters. On the other hand, the extended model reproduces the simulations very well, as illustrated by the solid blue lines in Fig. \ref{fig4} and also in Fig. \ref{fig5}.

The fact that $k_0$ clearly exceeds unity is also worth noting. As was briefly mentioned in the previous section, this might indicate the presence of additional internal structure on the strings--- commonly called wiggles. We leave a detailed study of this possible small-scale structure for future work---see \cite{Hindmarsh:2008dw} for early attempts to address this issue in field theory simulations, \cite{FRAC} for a similar analysis in Goto-Nambu simulations, and \cite{wiggly1,wiggly2} for extensions of the VOS model which explicitly account for small-scale structure.

It is especially instructive to compare the calibrated model parameters for the cosmic strings and domain walls cases. First, the normalization parameters, $k_0$ and $q$, are clearly different in the two cases, being larger for domain walls. This is not surprising; on the contrary, it makes sense that they depend on the dimensionality of the defects. According to the definition of the parameter $q$, its measured values lead us to infer that for cosmic strings the maximal value of the speed, for which the momentum parameter would vanish, is approximately $v\sim0.66$, while for domain walls it is $v\sim0.50$ if only simulations with relativistic speeds are considered, or $v\sim0.55$ if all simulations (including non-relativistic and ultra-relativistic ones) are included.

Then we have the exponent parameters, $r$ and $\beta$; the former also differs, in this case being somewhat larger for cosmic strings, while for the latter the situation is less clear. Indeed, if one considers only domain walls with intermediate expansion rates (comparable to those that we have explored, for cosmic strings, in the present work), then one finds that a value around $\beta=1.5$ is preferred both for cosmic strings and domain walls. However, when a broader range of expansion rates is considered in the analysis (including both ultra-relativistic and non-relativistic networks) then the domain walls VOS model prefers a value around $\beta=1.0$. It will be interesting to also study the non-relativistic case for cosmic strings, which we leave for subsequent work (as previously mentioned, our current discretization algorithm does not straightforwardly deal with very fast expansion rates).

Even more interesting---as well as more important given the potential observational implications---is the behavior of the energy loss parameters, $c$ and $d$. The latter (quantifying the losses due to scalar radiation) is the parameter that varies the least between the four cases, and it is tempting to speculate that there should be a universal value for it, applicable to all topological defects and presumably with a value around $d=0.26$; it will be interesting to test this hypothesis in field theory simulations of other defects, such as monopoles \citep{Monopoles} or semilocal strings \citep{Semilocals}.

Last but not least, we come to the loop chopping efficiency $c$, for which the difference between cosmic strings and domain walls is most striking. For domain walls we consistently find $c=0$, implying that scalar radiation is sufficient to explain the energy losses seen in wall simulations, while the production of wall blobs (the analogues of cosmic string loops) is not dynamically significant. On the other hand, for cosmic strings we find a value of $c$ that is not only clearly different from zero (at a very high level of statistical significance) but indeed somewhat higher than $d$. This clearly shows that the physical processes underlying the energy loss mechanisms are different in both cases. We will return to this point in the conclusions.

Since the two parameters multiply different functions of velocity, a more instructive comparison comes from evaluating the ratio of the two energy loss terms in the evolution equation for the correlation length. (Note that the two terms will also appear, in the same proportion, in the corresponding evolution equation for the energy density of the string network.) We define this ratio as
\begin{equation}
\Omega=\frac{\text{Loop losses}}{\text{Radiation losses}}=\frac{cv}{d[k_0-k(v)]^r}\,.
\end{equation}
Using the fitted model parameters and the velocities directly measured from the simulations we find that in the radiation era ($m=1/2$) the ratio is
\begin{equation}
\Omega_{rad}\sim0.82\,
\end{equation}
while in the matter era ($m=2/3$) it is
\begin{equation}
\Omega_{mat}\sim1.06\,;
\end{equation}
so in the latter era loop production should (marginally) dominate while in the radiation era scalar and gauge radiation is more important. Note that this confirms the previously stated expectation that scalar radiation losses should be less important for faster expansion rates, which correspond to smaller velocities. As a final comparison, if we take the faster expansion rate $m=0.9$ the ratio has the value
\begin{equation}
\Omega_{0.9}\sim6.92\,,
\end{equation}
and therefore radiation is completely subdominant.

As a final remark, it is instructive to compare the newly determined value of the loop chopping efficiency $c$ to those found for the previous version of the VOS model. The work of \cite{Moore:2001px,Moore} found $c=0.23\pm0.04$ from a comparison of the model to Goto-Nambu simulations in the radiation and matter eras, and $c=0.57\pm0.05$ from a comparison of the model to field theory simulations in the radiation and matter eras (and also to Goto-Nambu simulations in Minkowski spacetime). Our new result differs from the former at the level of two standard deviations, a difference which is quite understandable given that this parameter is somewhat correlated with other model parameters, and in particular with $k(v)$ of which the form has also changed in the extended model. On the other hand the new result is significantly smaller than the previous value obtained from field theory simulations, which is again to be expected given that we now have an additional radiation term accounting for some of the network's energy losses.

\section{\label{concl}Conclusions}

In this work we took advantage of our recently developed field theory cosmic string evolution code for the $U(1)$ model, which exploits the Compute Unified Device Architecture such that it uses GPUs as accelerators, and has been previously validated in \cite{Correia:2018gew}), to extend and accurately calibrate the VOS model. The code speed has enabled us to simulate an extensive set of cosmic string networks in expanding universes for 43 different expansion rates in an extremely comfortable amount of time---about one day's work for each our two production runs. Indeed the data analysis of the simulation diagnostics took far longer than running the simulations themselves.

This large number of different expansion rates is crucial for an accurate model calibration, since it makes it possible to infer the detailed velocity dependence of the relevant physical mechanisms encoded in the VOS model, thereby breaking degeneracies that would otherwise exist between the various model parameters (which are now six rather than two). The relevance of exploring this dimension of parameter space was already exhibited in previous work on domain walls \cite{Rybak1,Rybak2}, and our present results confirm its importance.

Our analysis shows that the energy loss mechanisms in the cosmic strings VOS model should be extended to account for radiation by fields, and that the previous analytic ansatz for the momentum parameter is inadequate to fully reproduce simulations in a wide range of cosmological settings (specifically, with various expansion rates). Our extensions to the VOS model lead to very satisfactory agreement throughout the simulated range of expansion rates. Importantly, we have found that unlike the domain walls case (in which scalar radiation can completely account for the energy losses), for strings the loop production and radiative loss terms are comparable, and indeed the former will dominate for fast enough expansion rates---roughly $m>0.65$, thus including the matter era.

In the future we will address the non-relativistic version of the momentum parameter, by simulating these networks in expansion rates larger than what was considered in this manuscript. A comparison with Goto-Nambu simulations---ideally over an equally extensive range of expansion rates---is also highly desirable, as a further test of this model. Note that in Goto-Nambu simulations the strings will not undergo losses due to radiation (in terms of modeling we effectively have $d\rightarrow0$), and that the form of the momentum parameter in the previous version of the VOS model was mainly inferred from Goto-Nambu simulations (see \cite{MS2} for a detailed discussion), so it is \textit{a priori} not obvious that the model as calibrated by field theory simulations will perform equally well for Goto-Nambu simulations. Such a comparison will therefore be an important test of the model.

Our work shows that there is a tangible performance benefit to using GPUs in field theory defect simulations, enabling the possibility of running thousands or tens of thousands of high-resolution simulations in quite acceptable amounts of time. This opens several interesting possibilities for the further exploration of the cosmological consequences of these networks. A long-term open issue in the cosmic strings literature is the apparent inconsistencies in the results obtained in Goto-Nambu simulations, for which there are several independent codes \citep{BB,AS,FRAC,VVO,Blanco}, and in field theory simulations, for which all recent results preceding our work ultimately came from one single code \cite{Bevis:2006mj,Hindmarsh:2017qff}.

We do emphasize that, to the extent that a comparison can be made, the numerical results of our code \cite{Correia:2018gew} are fully consistent with those of \cite{Bevis:2006mj,Hindmarsh:2017qff}, though our interpretation of them, which is illuminated by the physical content of the VOS model, is slightly different. One point to bear in mind, regarding the perceived differences between the results of Goto-Nambu and field theory simulations of cosmic strings is that both the physical content and the numerical diagnostics differ in subtle ways. In the Goto-Nambu case one has a loop production function which is (at least in principle) well defined and easy to extract from simulations, although it is sometimes confused with the loop distribution function (including all loops present in the simulation box at a given moment, and not just the ones recently produced). For field theory simulations one has a generalized energy loss function which is scale-dependent in a non-trivial way. This function will have contributions from loop production on correlation length scales (typically one such loop being produced per Hubble volume per Hubble time), but also from the production of so-called proto-loops and blobs on scales around the defect thickness, and from scalar radiation on a wide range of scales. In this latter context what one decides to call a loop in the usual (Goto-Nambu) sense is, to some extent, a matter of choice. We leave a more detailed study of these different mechansims for subsequent work, while emphasizing that from the point of view of observational consequences the important diagnostic is the overall energy loss function, and this can still be measured unambiguously in both types of simulations. In any case, we note that the availability of an improved (better calibrated) VOS model can itself enable a more detailed and quantitative comparison between the results of the two types of codes, and shed light on these apparent inconsistencies.

Our calibration was done relying on a range of constant expansion rates $m$, which are simpler both from the computational and the post-processing points of view, the reason for the latter being that the networks are expected to reach scaling (as is indeed confirmed by our work) for all constant values of $0<m<1$. In this work we focused on the range of expansion rates leading to relativistic scaling networks, $0.5\le m\le0.95$; we leave the study of the non-relativistic ($m>0.95$) and ultra-relativistic ($m<0.5$) cases for subsequent work. It will also be interesting to verify if the VOS model thus calibrated can accurately reproduce the cosmological radiation-to-matter transition (which has been shown to be the case for the analogous domain wall model \cite{Rybak1}) and also the matter-to-acceleration transition.

In the longer term, an optimized multi-GPU code (which is currently under active development) can be used to produce thousands or even tens of thousands of accurate full-sky maps of cosmic microwave or gravitational wave backgrounds, which can be used in the data analysis pipelines of ongoing as well as next-generation experiments, including CORE \cite{CORE} or LISA \cite{LISA}. This will eliminate the current bottleneck in these analyses (so far one can only generate a few full-sky maps, or alternatively many maps of very small sky patches, with a reliable resolution) and will therefore lead to more robust as well as more stringent constraints on topological defects, cosmological phase transitions and GUTs. We do expect that GPU-based defect codes will in the medium term become the gold standard in the field. In this new era of GPU-based defect simulations, the role of the VOS and other such analytic models has also changed.

\begin{acknowledgments}

This work was financed by FEDER---Fundo Europeu de Desenvolvimento Regional funds through the COMPETE 2020---Operational Programme for Competitiveness and Internationalisation (POCI), and by Portuguese funds through FCT - Funda\c c\~ao para a Ci\^encia e a Tecnologia in the framework of the project POCI-01-0145-FEDER-028987. J.R.C. is supported by an FCT fellowship (SFRH/BD/130445/2017). We gratefully acknowledge the support of NVIDIA Corporation with the donation of the Quadro P5000 GPU used for this research.

Useful conversations on the topic of this work with Ana Ach\'ucarro, Tasos Avgoustidis, Asier Lopez-Eiguren, Ivan Rybak, Patrick Peter and Paul Shellard are gratefully acknowledged.

\end{acknowledgments}

\bibliography{artigo}

\end{document}